\documentclass{ieeeaccess}
\usepackage{cite}
\usepackage{amsmath,amssymb,amsfonts}
\usepackage{algorithmic}
\usepackage{graphicx}
\usepackage{textcomp}
\usepackage{gensymb}
\usepackage[caption=false,font=footnotesize]{subfig}

\def\BibTeX{{\rm B\kern-.05em{\sc i\kern-.025em b}\kern-.08em
    T\kern-.1667em\lower.7ex\hbox{E}\kern-.125emX}}
\begin{document}
\history{Date of publication xxxx 00, 0000, date of current version xxxx 00, 0000.}
\doi{10.1109/ACCESS.2017.DOI}
\newlength{\xfigwd}
\setlength{\xfigwd}{\columnwidth}

\title{Frequency \& Radiative Analysis of Random Yagi-UHF/VHF Phased Array}
\author{\uppercase{Luis M. Bres}\authorrefmark{1},
\uppercase{Luis A., Hernandez}\authorrefmark{2}, and \uppercase{Teviet D. Creighton}\authorrefmark{3},
\IEEEmembership{Member, IEEE}}
\address[1]{Department of Physics and Astronomy, South Texas Space Science Institute, University of Texas Rio Grande Valley, TX 78520 USA (e-mail: luismario.brescastro01@utrgv.edu)}
\address[2]{Department of Physics and Astronomy, South Texas Space Science Institute, University of Texas Rio Grande Valley, TX 78520 USA (e-mail: luis.hernandez16@utrgv.edu)}
\address[3]{Department of Physics and Astronomy, South Texas Space Science Institute, University of Texas Rio Grande Valley, TX 78520 USA (e-mail: teviet.creighton@utrgv.edu)}
\tfootnote{This project was supported by funding from the University of Texas Rio Grande Valley.}

\markboth
{Author \headeretal: Preparation of Papers for IEEE TRANSACTIONS and JOURNALS}
{Author \headeretal: Preparation of Papers for IEEE TRANSACTIONS and JOURNALS}

\corresp{Corresponding author: Luis M. Bres (e-mail: luis.bres95@gmail.com).}

\begin{abstract}
This paper investigates a phased array ground station capable of tracking multiple sources, multi-beamforming, electronic steering, easy scaling, and low cost. The project will develop a 20-pair dual-polarized yagi-UHF/VHF phased array with a pseudo-random layout, comparing parameters of random and uniform distributions. We will present several analyses: general analysis for side lobes across both elevation and azimuth, analysis of scaling with number of elements (``element sweep''), electronic beam steering analysis, mechanical beam steering analysis, electro-mechanical beam steering analysis, array density analysis, and reception/transmission spectra analysis.
\end{abstract}


\titlepgskip=-15pt

\maketitle

\section{Introduction}
\label{sec:introduction}
\PARstart{T}{he} cost of launching payloads to Low-Earth Orbit (LEO) has dropped significantly~\cite{b1}, driving a surge in satellite deployments and increasing the need for efficient, multi-track ground stations~\cite{b2}.

Traditional parabolic antennas provide strong beamforming but are limited to single-target tracking, require mechanical movement, and lack electronic beamforming~\cite{stutzman2013}. In contrast, phased arrays manipulate signal amplitudes and phases to enhance the main lobe, reduce side lobes, enable electronic steering, improve signal-to-noise ratio, and support multiple links
~\cite{balanis}.

A planar, pseudo-random array was chosen for its superior side lobe reduction compared to uniform distributions~\cite{b5}. The selected UHF and VHF frequencies align with CubeSat traffic, where 51\% of 120 missions used UHF and 7\% used VHF, making them ideal for our ground station.

Random arrays have been explored across a variety of applications, including medical imaging~\cite{chaplin2018,zubair2021}, general validation studies~\cite{8597878}, and performance enhancement through random sub-array configurations~\cite{valle2024}. Early investigations into random linear arrays~\cite{1147413} demonstrated their capability for effective side lobe attenuation. More recently, randomness in array configurations has been examined in the context of Internet of Things (IoT) systems~\cite{wang2018}, where element position uncertainties were shown to significantly impact phased array performance, providing practical design insights.

In contrast, phased arrays for satellite communications are predominantly implemented as uniform panel arrays composed of patch antennas. In such systems, side lobe reduction is typically achieved by increasing the number of elements~\cite{9515267,chen2022}, which leads to higher system complexity and cost, particularly in the associated back-end hardware.

However, despite these advancements, prior work generally evaluates side lobe attenuation using only a E-plane cut. As a result, the full spatial behavior of side lobes across both H-plane plane remains insufficiently characterized.

\section{Metrics}
\subsubsection{Axes}
We will follow common conventions of coordinate axes. The electric polarization is aligned with the elevation $\theta$ plane (E-plane), and the magnetic polarization is aligned with the azimuth $\phi$ plane (H-plane).  The array elements are always arranged in a horizontal plane, with $\theta=90^\circ$ representing the zenith direction normal to that plane. 

\begin{figure}[t!]
    \centering
    \includegraphics[width=0.9\columnwidth,clip=true]{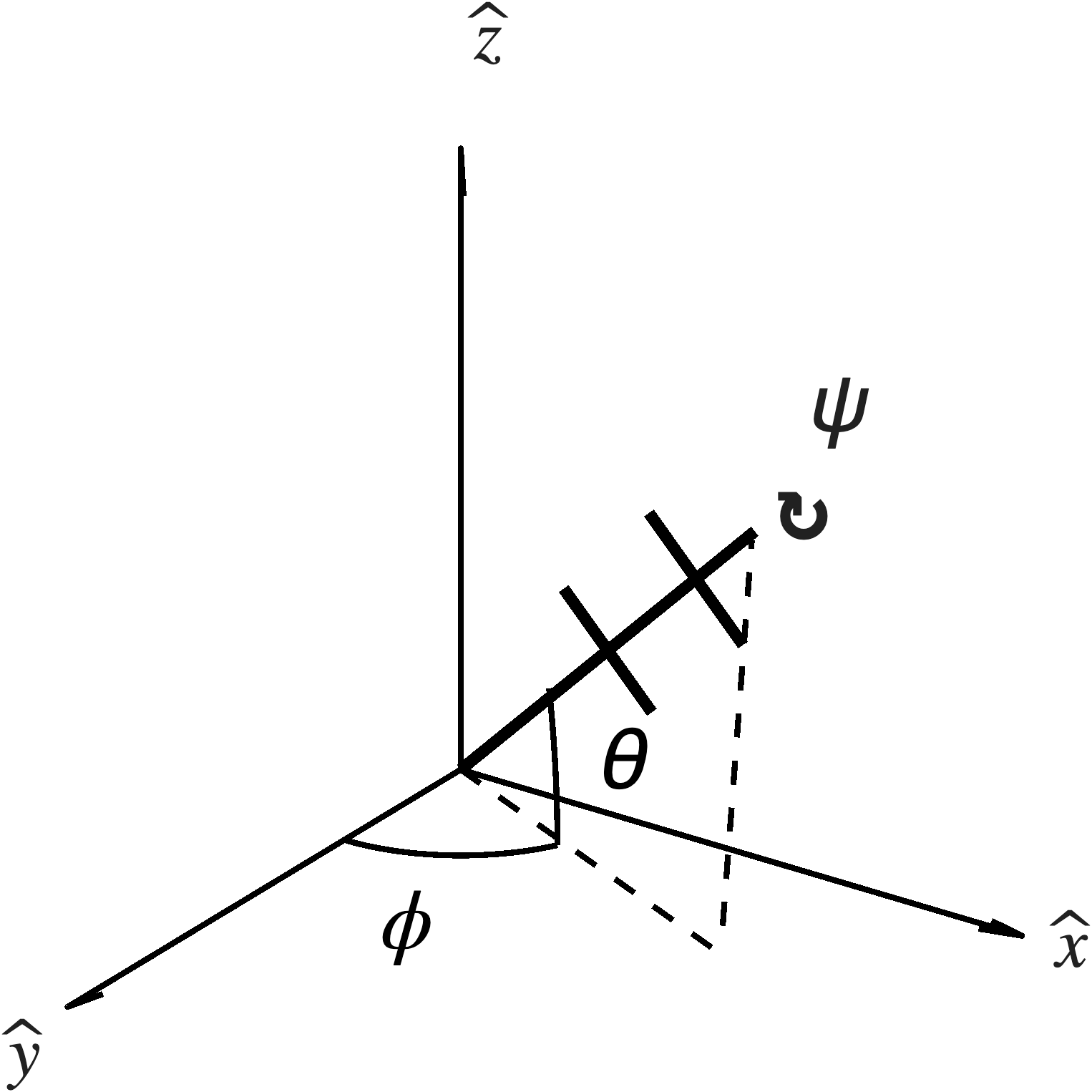}
    \caption{Antenna's axis framework.}
    \label{AxisFig}
\end{figure}

Fig.~\ref{AxisFig} shows the antenna's axis framework, where we previously defined the $\theta$ and $\phi$ angles of rotation for the E-plane and H-plane respectively. We introduce a third axis degree of freedom $\psi$, which describes the antenna's own directional axis of rotation.  (For $\psi\neq0$ the ``elevation'' E-plane will no longer align with the electric field, and the ``horizontal'' H-plane will not align with the magnetic polatization.)  This will be relevant on the study cases for section~\ref{Frequency}.

\subsubsection{Isotropic Gain [dBi]}
Isotropic gain quantifies the logarithmic power ratio between an antenna gain ($P_\mathrm{ant}=P_\mathrm{rad}/P_\mathrm{input}$) and an isotropic antenna gain ($P_\mathrm{iso}$). 

\begin{equation}
    G_\mathrm{dBi}=10\log_{10}\left(\frac{P_\mathrm{ant}}{P_\mathrm{iso}}\right)
\end{equation}

\subsubsection{Beam Pattern [dBi]}
The beam pattern is the radiation emitted from the antenna as a function of space coordinates and frequency. The radiation field that the antenna emits also indicates its sensitivity to received signals. A non-resonant-frequency beam pattern will be less efficient and powerful than a resonant-frequency beam pattern, hence the importance of the reception spectra~\cite{vantrees2002}.

The yagi-UHF beam pattern ($f_r=435$ MHz) has a $\mathbf{ML}$ of 13 dBi. Fig.~\ref{UHFbp1v20}a shows that the short spacings, short length, and high number of dipole elements of the antenna will produce a sharp beam pattern.

The yagi-VHF beam pattern ($f_r=145$ MHz) has a $\mathbf{ML}$ of 9.44 dBi. Fig.~\ref{UHFbp1v20}c shows that the long spacings, long length, and low number of dipole elements of the antenna will produce a broad beam pattern.

\subsubsection{Array Factor}
To calculate the beam pattern of a phased array one must compute the array factor and multiply it by the beam pattern of an individual antenna~\cite{mailloux2005}. The array factor is a complex value representing the amplitude and phase of the signal at the array elements for a given direction.

\begin{equation}
AF_{\phi,\theta}=\sum_{n=1}^{N}w_ne^{i\left(\vec{k}(\phi,\theta)-\vec k_s\right)\cdot \vec r_n}\label{AF}
\end{equation}

\begin{equation}
w_n=A_ne^{-i\vec k_s\cdot \vec r_i}
\end{equation}

The subscripts of $\phi$ and $\theta$ are the angles of the array factor where the spherical wave vector $\vec{k}(\phi,\theta)$ is being scanned. The steering wave vector $\vec k_s$, in spherical terms as well, represents phase shifts applied to antenna signals before combining, and allows for electronic steering of the array towards a specified azimuth and elevation. The symbol $k$ is the wave number. The parameter $w_n$ represents the antenna element complex weight which includes its amplitude, $A_n$ and its phase shift component. Under normal conditions and for basic steering strategies all amplitudes $A_n$ will be set to 1. Amplitude weight tapering strategies can be used to attenuate side lobe magnitudes at the expense of increasing the main lobe full-width half-maximum. 
\begin{equation}
 \vec{k}(\phi,\theta)=|k|(\hat{x}\cos\theta\cos\phi+\hat{y}\cos\theta\sin\phi+\hat{z}\sin\theta)
\end{equation}

\begin{equation}
 \vec k_s=|k_s|(\hat{x}\cos\theta_s\cos\phi_s+\hat{y}\cos\theta_s\sin\phi_s+\hat{z}\sin\theta_s)
\end{equation}

To get the final phased array beam pattern $F(\phi,\theta)$ we need to calculate the array factor's squared magnitude, normalize it by the maximum of the array factor's magnitude, and then we multiply it by the antenna individual pattern in linear scale $P_\mathrm{ant,lin}(\phi,\theta)$ and finalize the process by converting it to dBi scale~\cite{johnson1993}. 

The array factor's squared magnitude defines the power pattern, which we normalize using the maximum of the array factor's magnitude to preserve the directivity peak. Normalizing against the power pattern's own maximum would distort the distribution and reduce apparent directivity by normalizing the peak to 1~\cite{app14135917}.
\begin{equation}
P_\mathrm{ant,lin}(\phi,\theta)=10^{P_\mathrm{ant,dBi}/10}
\end{equation}
\begin{equation}
F(\phi,\theta)_\mathrm{dBi} = 10 \log_{10} \left( P_\mathrm{ant,lin}(\phi,\theta) \frac{\lvert AF_{\phi,\theta} \rvert^2}{\lvert AF_{\phi,\theta} \rvert_\mathrm{max}} \right)
\end{equation}
\subsubsection{Main Lobe and Side Lobe [dBi]}

\begin{figure}[t!]
\centering
\begin{minipage}{0.48\columnwidth}
    \centering
    \includegraphics[width=\columnwidth,clip=true]{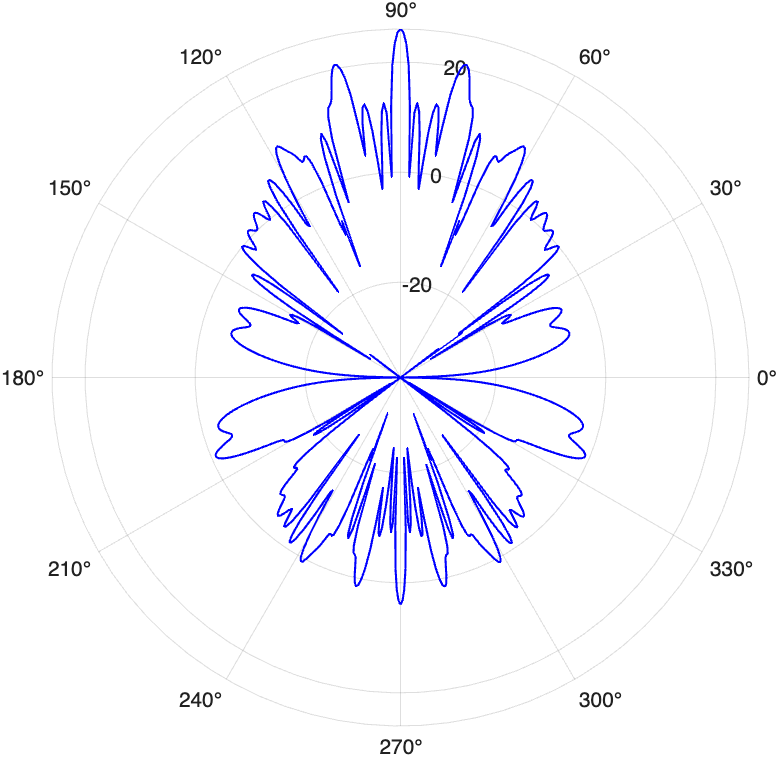}
    
\end{minipage}\hfill
\begin{minipage}{0.48\columnwidth}
    \centering
    \includegraphics[width=\columnwidth,clip=true]{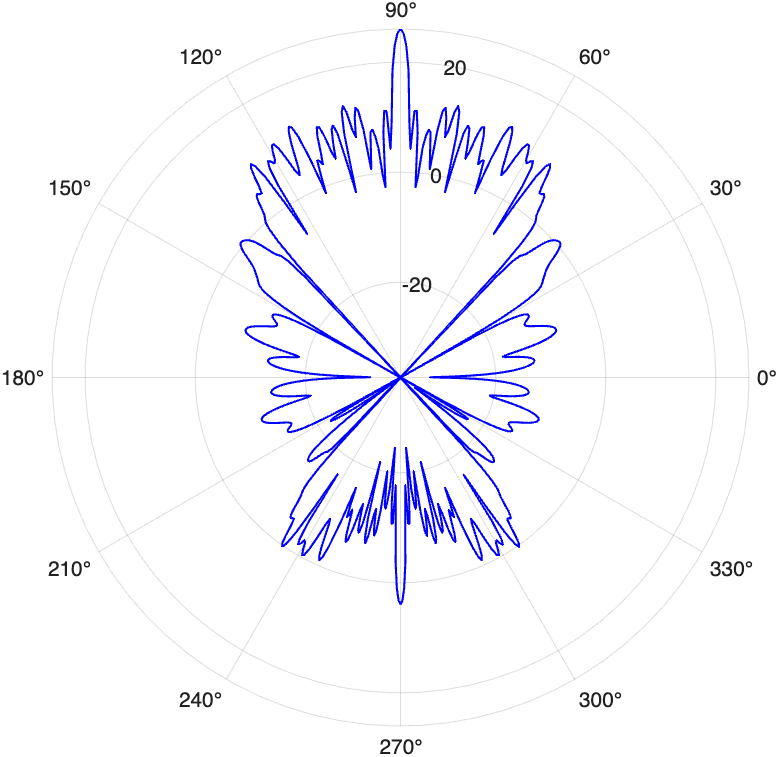}
\end{minipage}
\caption{Yagi-UHF random array E-plane beam pattern cuts evaluated at $-98^\circ$ and $-16^\circ$ H-plane angles.}
\label{bpslices2D}
\end{figure}

The main lobe $\mathbf{ML}$ of a beam pattern is the field region with the highest intensity, characterized by its magnitude, beam width, and direction, while the side lobe is the next peak intensity region, usually surrounding the $\mathbf{ML}$~\cite{antennaTheory}. We want a narrow, maximized beam of the $\mathbf{ML}$ and attenuated side lobes. We will define the peak side lobe value of a E-plane ($\theta$) beam pattern at any given H-plane angle $\phi$ as the side lobe angular magnitude, $\mathbf{SL(\phi)}$.
Fig.~\ref{bpslices2D} demonstrates the variability of $\mathbf{SL(\phi)}$, indicating that a single E-plane slice is insufficient to characterize the beam pattern, displaying two beam pattern E-plane cuts at different H-plane angles to highlight side lobe level differences.

\subsubsection{Half-Power Beam-Width, $BW_{HP}$,[$\degree$]}
$\mathbf{BW_{HP}}$ is the angular beam-width of the E-plane main lobe's half power.

\begin{figure}[b!]
    \centering
    \begin{minipage}[t]{0.4\columnwidth}
        \centering
        \includegraphics[height=0.15\textheight]{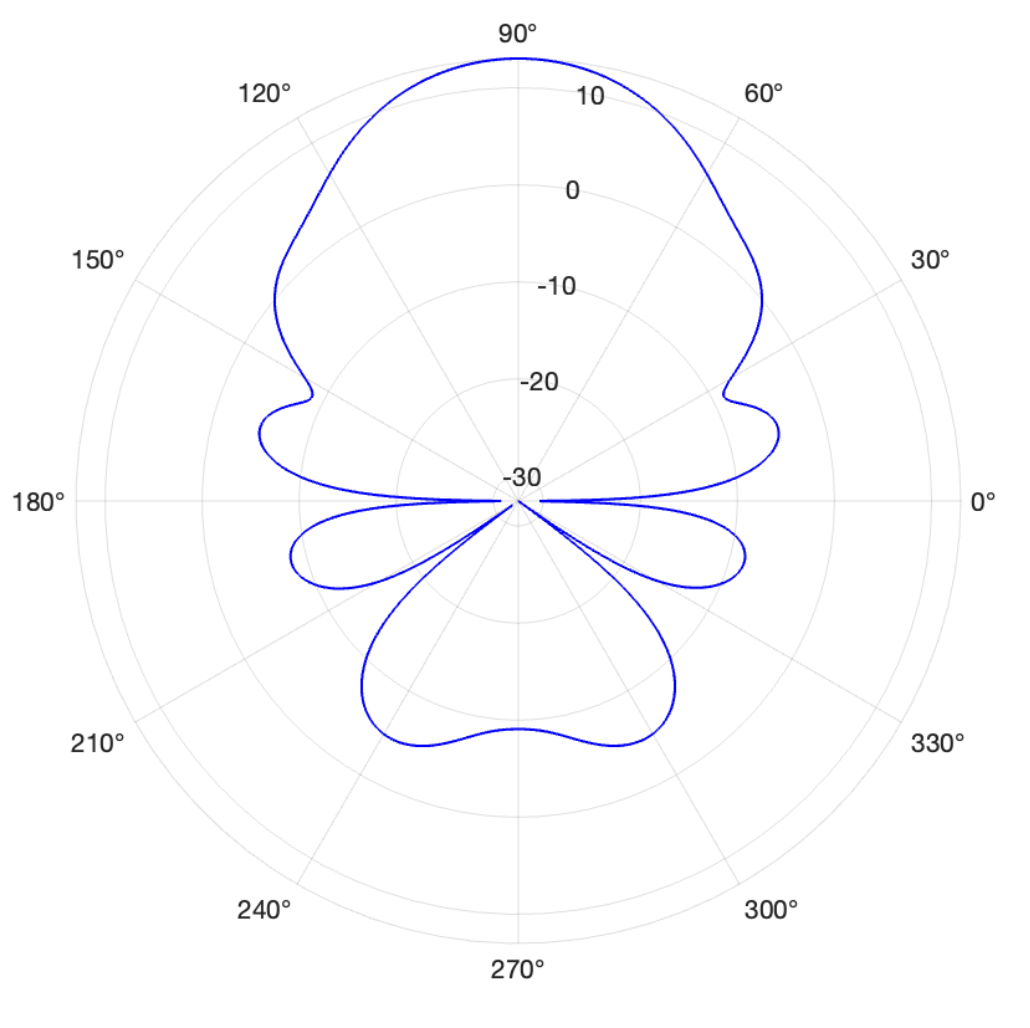}
        \vspace{1ex}
        
        \small a) Single Yagi--UHF antenna.
    \end{minipage}\hfil
    \begin{minipage}[t]{0.4\columnwidth}
        \centering
        \includegraphics[height=0.15\textheight]{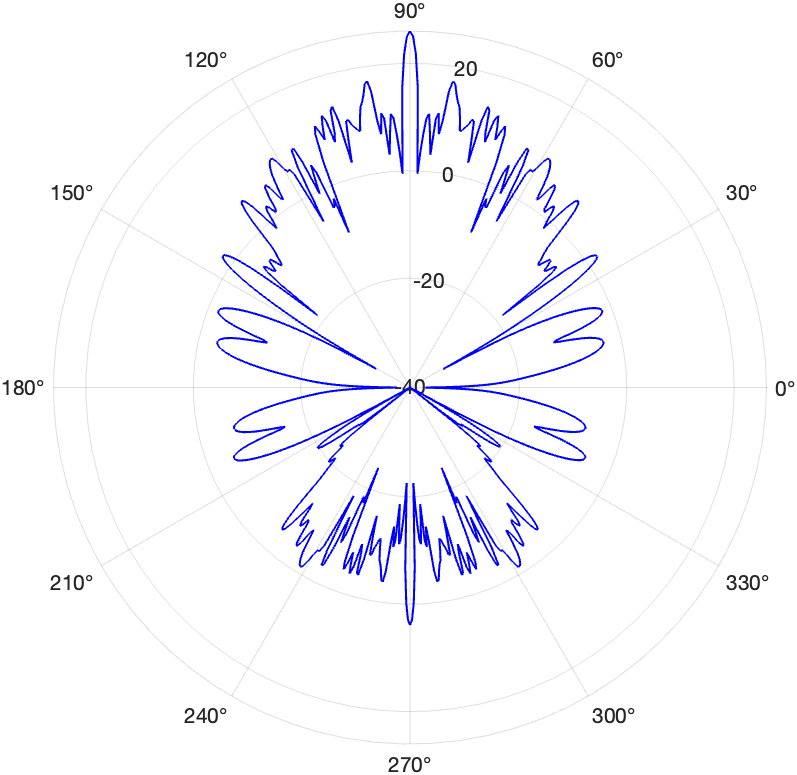}
        \vspace{1ex}
        
        \small b) Yagi--UHF random array.
    \end{minipage}

    \vspace{2ex}

    \begin{minipage}[t]{0.4\columnwidth}
        \centering
        \includegraphics[height=0.15\textheight]{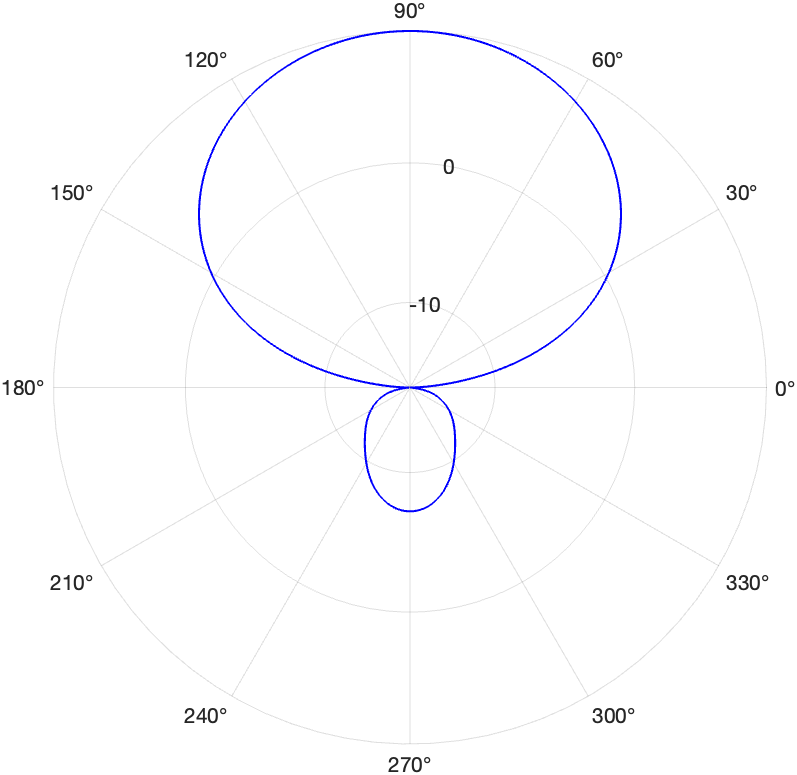}
        \vspace{1ex}
        
        \small c) Single Yagi--VHF antenna.
    \end{minipage}\hfil
    \begin{minipage}[t]{0.4\columnwidth}
        \centering
        \includegraphics[height=0.15\textheight]{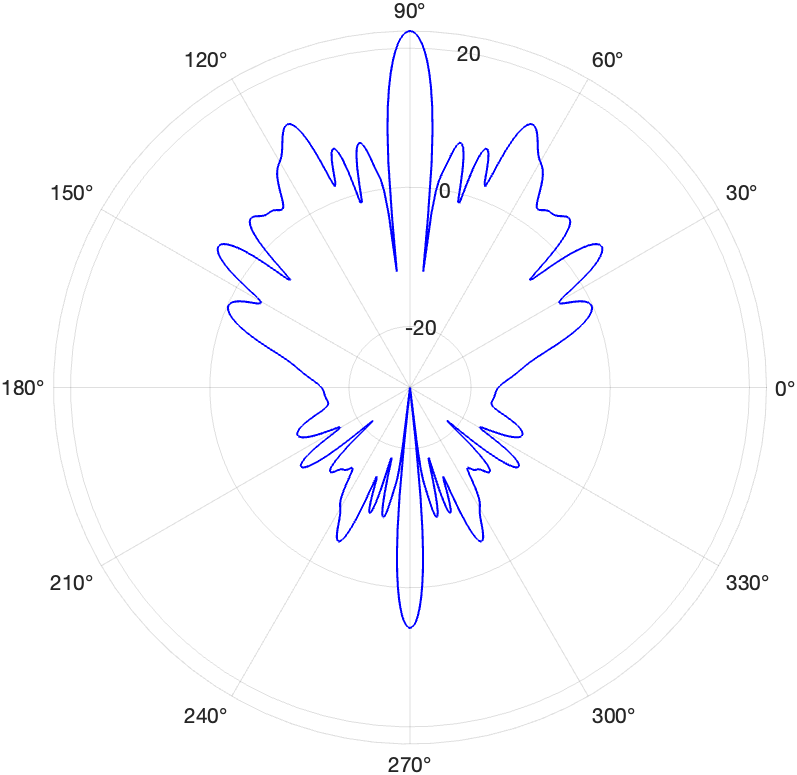}
        \vspace{1ex}
        
        \small d) Yagi--VHF random array.
    \end{minipage}

    \caption{E-plane beam patterns cuts for single yagi-UHF/VHF antennas and yagi-UHF/VHF random arrays.}
    \label{UHFbp1v20}
\end{figure}

\subsubsection{Main Lobe to Clutter Ratio, $MLC$ [dB]}
The $\mathbf{MLC}$ ratio is the pattern's main lobe $\mathbf{ML}$ peak value over the beam pattern average, $\mathbf{\langle BP \rangle}$. We are treating the $\mathbf{ML}$ as a target source, therefore the target's solid angle should be infinitesimal. If the $\mathbf{MLC}$ ratio is greater than zero, then $\mathbf{ML}$ is multiples of $\mathbf{\langle BP \rangle}$. If the $\mathbf{MLC}$ ratio is less than zero, then $\mathbf{\langle BP \rangle}$ is multiples of $\mathbf{ML}$. This metric is in dB because it is the ratio between $\mathbf{ML}$ and $\mathbf{\langle BP \rangle}$.
\begin{equation}
MLC = \frac{ B_{peak} }{ \langle B \rangle_{\Omega_{\text{BP}}} }
= \frac{
\displaystyle B_{peak}(\theta_{ML}, \phi_{ML})
}{
\displaystyle \frac{1}{\Omega_{\text{BP}}} \int_{\Omega_{\text{BP}}} B(\theta, \phi)\, d\Omega
}
\end{equation}
where the total solid angle of a rectangular region $R$ in spherical coordinates is the following:
\begin{equation}
    \Omega_{BP} = \int_{\Omega_{BP}} d\Omega = \int_{\phi_1}^{\phi_2} \int_{\theta_1}^{\theta_2} \cos\theta \, d\theta \, d\phi
\end{equation}

\subsubsection{Reception \& Transmission Spectra, [dB]}
The reception spectra of an antenna is a plot of the power absorbed as a function of frequency. A value of $0$ dB on the reception spectrum implies most power is reflected, while values significantly below zero indicate power absorption at that frequency~\cite{bird2009}. Transmission spectra are plots of power transferred from antenna 1 to antenna 2; $0$ dB values mean strong transmission, while values  below the resonant dip of the reception spectrum indicate low transmission at that frequency between antennas,~\cite{mathworks}.

\subsubsection{Field of view, $\Delta\theta$ [$\degree$]}

\begin{figure}[b!]
    \centering
    
    \begin{minipage}[t]{0.9\columnwidth}
        \centering
        \includegraphics[width=\columnwidth]{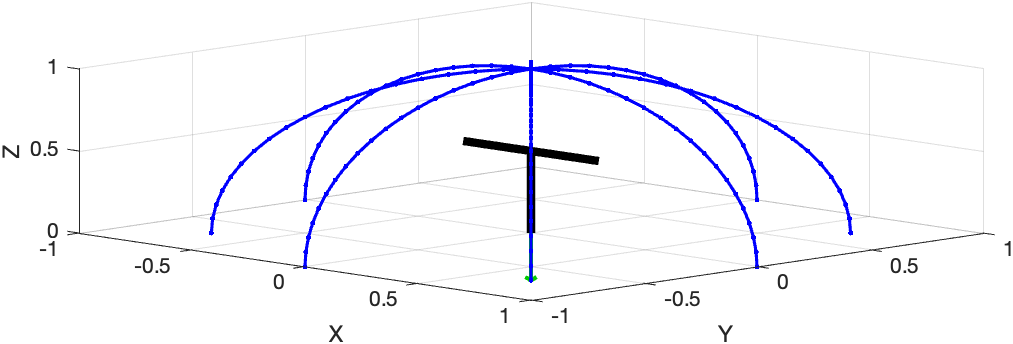}
        \vspace{1ex}

        \small a) Illustration of H-plane sweeps used to track beam pattern metrics.
    \end{minipage}
    
    \vspace{2ex}

    \begin{minipage}[t]{0.9\columnwidth}
        \centering
        \includegraphics[width=\columnwidth]{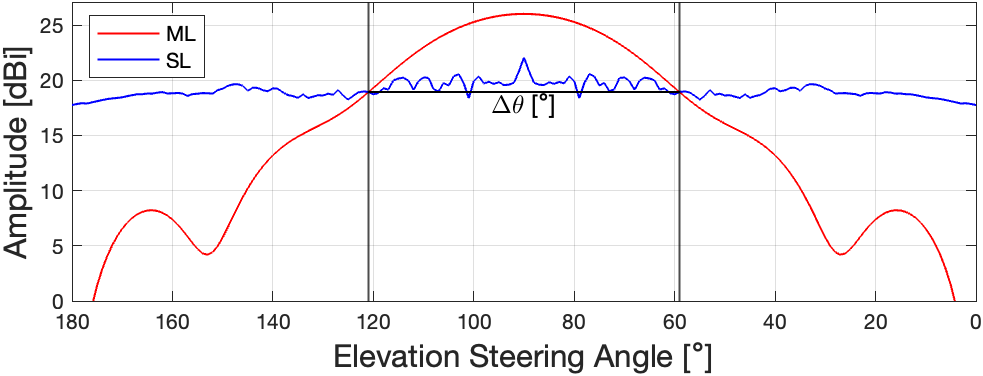}
        \vspace{1ex}

        \small b) Main lobe (ML) and highest side lobe (SL) tracking as a function of E-plane beam steering $\theta$ angle over a fixed H-plane $\phi$ angle.
    \end{minipage}

    \caption{Demonstration of the calculation of effective sky angle view $\Delta\theta$}
    \label{DthetaDemo}
\end{figure}

Fig.~\ref{DthetaDemo}a and~\ref{DthetaDemo}b explains the field of view angle in which the main lobe is equal to or higher than the highest side lobe. Fig.~\ref{DthetaDemo}a shows the E-plane beam steering sweeps across the H-plane. We will assess the fluctuations of metric $\Delta\theta$ as we would hope to be invariant across the H-plane. On Fig.~\ref{DthetaDemo}b, we can see how by looking at the full E-plane range the intersection points between the main lobe and side lobes, this represents the field of view angle $\Delta\theta$. This electronic beam-steering analysis was done with an E-plane sweep across a single H-plane angle.

\section{Frequency Analysis of Yagi-UHF/VHF Antennas}\label{Frequency}
We will study the effects of transmission spectra effects on our antennas by applying several test cases~\cite{incompliance}:

\begin{figure}[t!]
    \centering
    \includegraphics[width=0.9\columnwidth,clip=true]{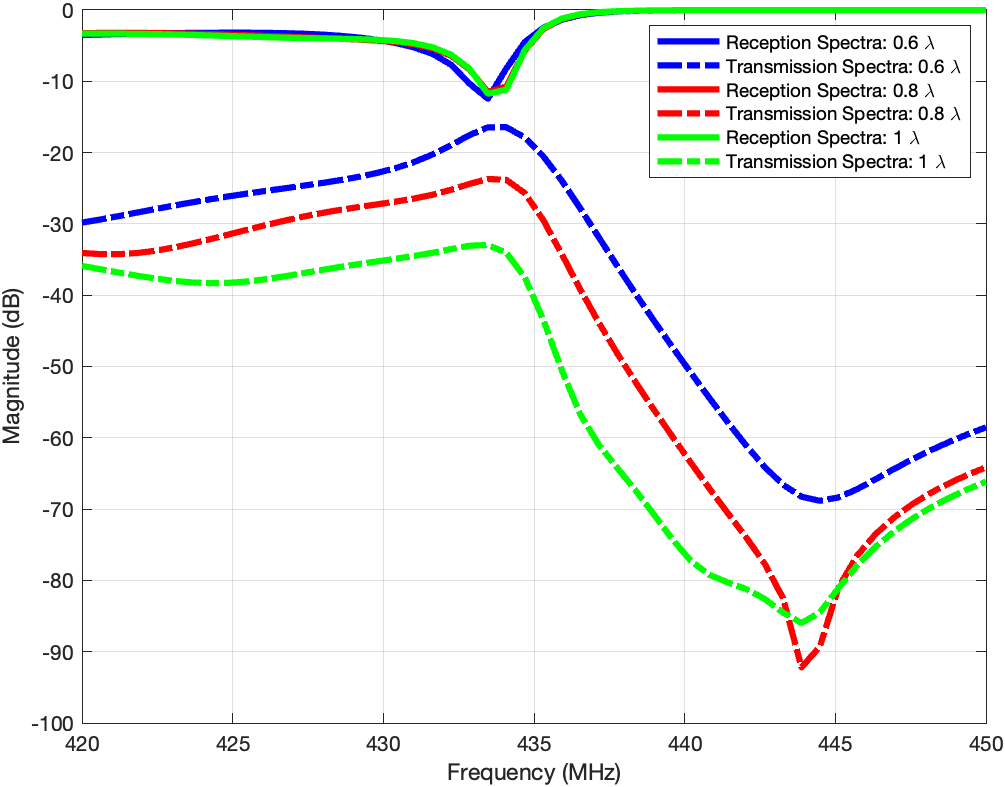}
    \caption{Case 1: Transmission spectra for various separation distances of a 2-element yagi-UHF antenna array.}
    \label{RT_distance_UHF}
\end{figure}

\begin{figure}[b!]
    \centering
    \includegraphics[width=0.9\columnwidth,clip=true]{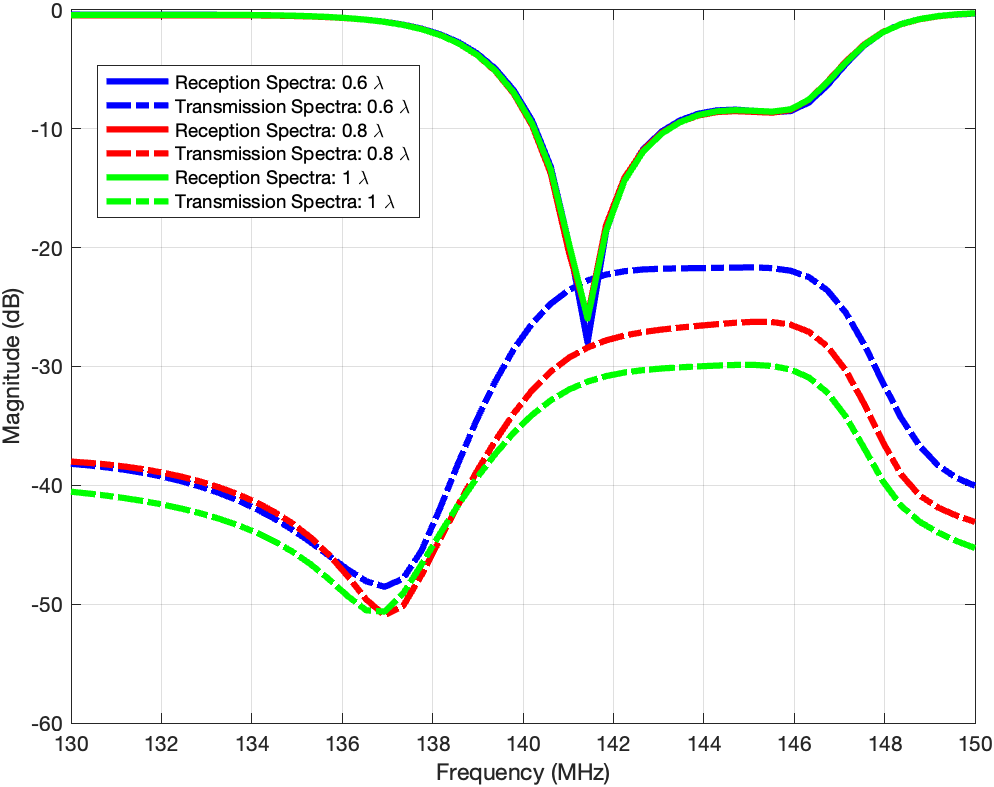}
    \caption{Case 1: Transmission spectra for various separation distances of a 2-element yagi-VHF antenna array.}
    \label{RT_distance_VHF}
\end{figure}

\subsubsection{Case 1: Separation distance sweep}
We will explore the effects of changing the separation distance between antennas. We define $\lambda_{UHF}=0.68$m and $\lambda_{VHF}=2.06$m as the resonant wavelengths of the yagi-UHF/VHF antennas. All antennas point at $90\degree$ in the E-plane. In Fig.~\ref{RT_distance_UHF} and Fig.~\ref{RT_distance_VHF} we observe how by increasing the separation distance between the antennas we lower the cross-element transmission interference. Fortunately for the UHF case, for none of the antenna separations did the transmission spectrum rise above the reception spectrum, indicating minimal cross-talk among antennas. For the VHF case, there is minimal transmission interference for separations greater than 1$\lambda_{VHF}$. We consider separations as low as $0.6\lambda_{UHF/VHF}$ due to mechanical constraints.
\begin{figure}[t!]
    \centering
    \includegraphics[width=0.9\columnwidth,clip=true]{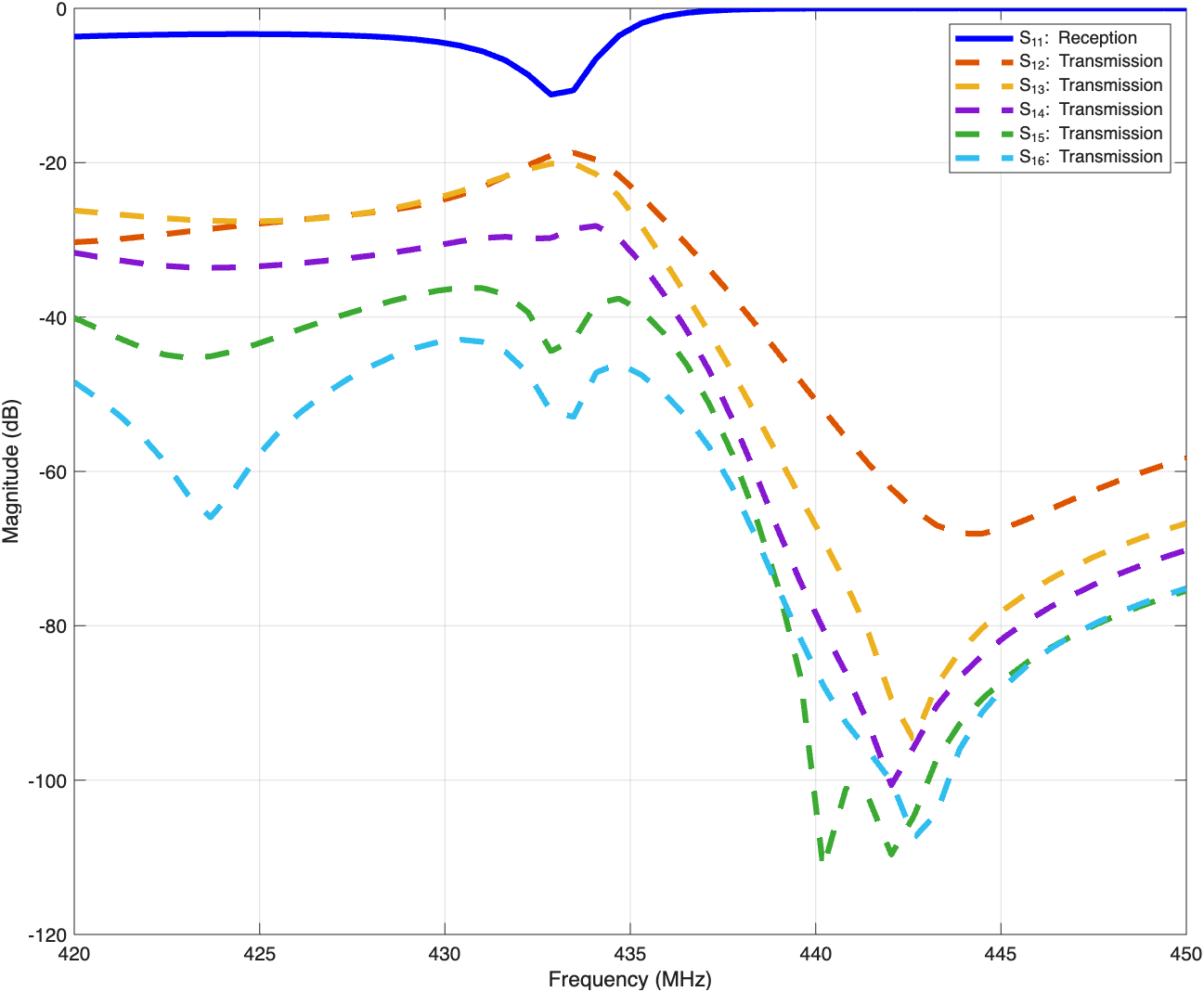}
    \caption{Case 2: Transmission spectra of a 6-element yagi-UHF antenna linear array.}
    \label{linearTspecUHF}
\end{figure}

\subsubsection{Case 2: Linear array reception/transmission spectra analysis}

\begin{figure}[b!]
    \centering
    \includegraphics[width=0.9\columnwidth,clip=true]{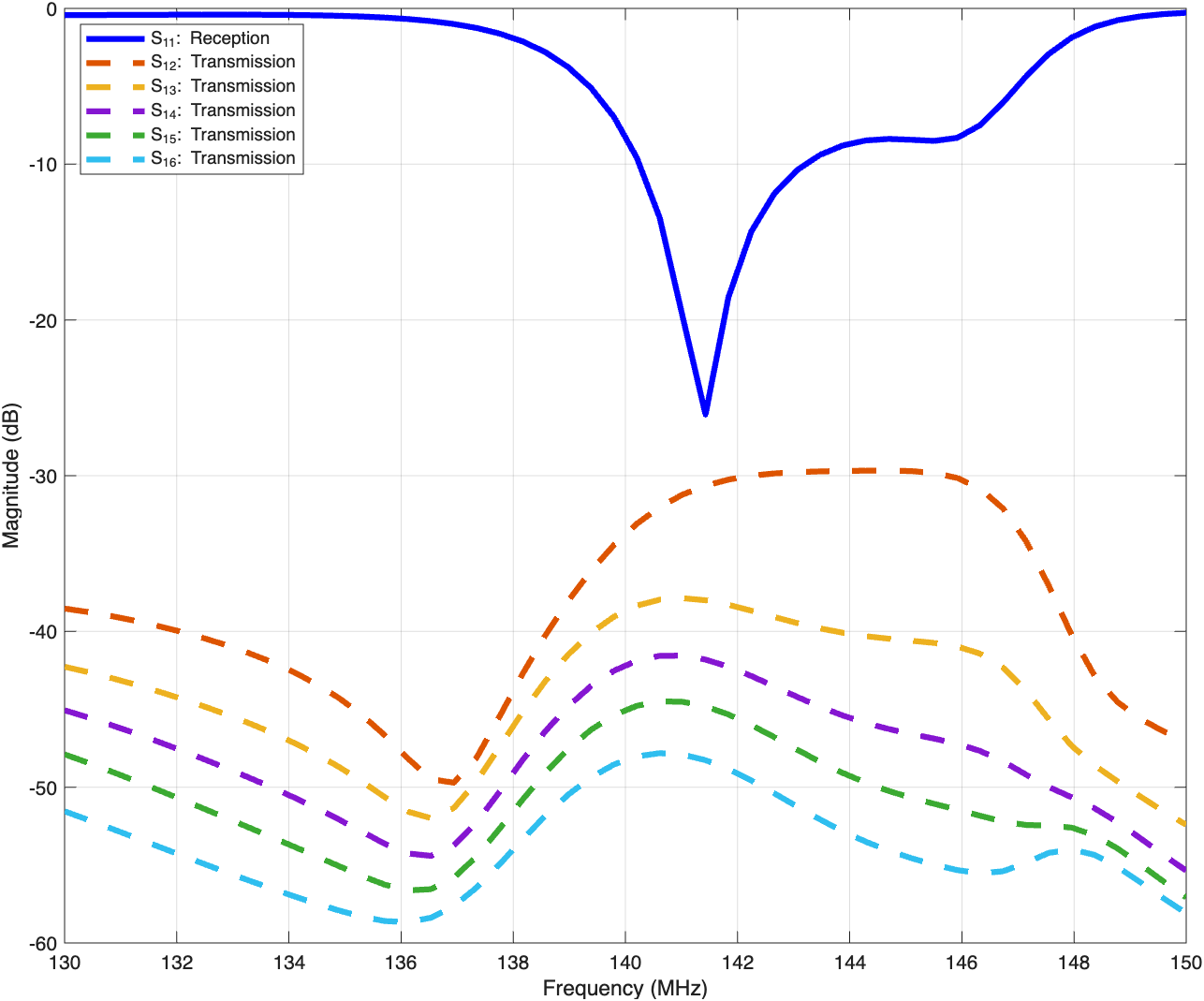}
    \caption{Case 2: Transmission spectra of a 6-element yagi-VHF antenna linear array.}
    \label{linearTspecVHF}
\end{figure}
Fig.~\ref{linearTspecUHF} and Fig.~\ref{linearTspecVHF} explore the higher-order effects on the antenna's transmission spectra including second-order reflections. The setup simulation will be a 6-element linear array with $0.6\lambda_{UHF/VHF}$ separation distance. We will evaluate the transmission spectra between the first antenna and the rest of the array.

We observe that for the UHF array, reflections follow an expected pattern where increasing distance lowers the transmission cross-talk:  there is no significant cross-talk (transmission exceeding reception) for any of the distances considered. The VHF case follows a similar trend, except there is non-negligible cross-talk for the first couple of separations. 


\begin{figure}[t!]
    \centering
    \includegraphics[width=0.9\columnwidth,clip=true]{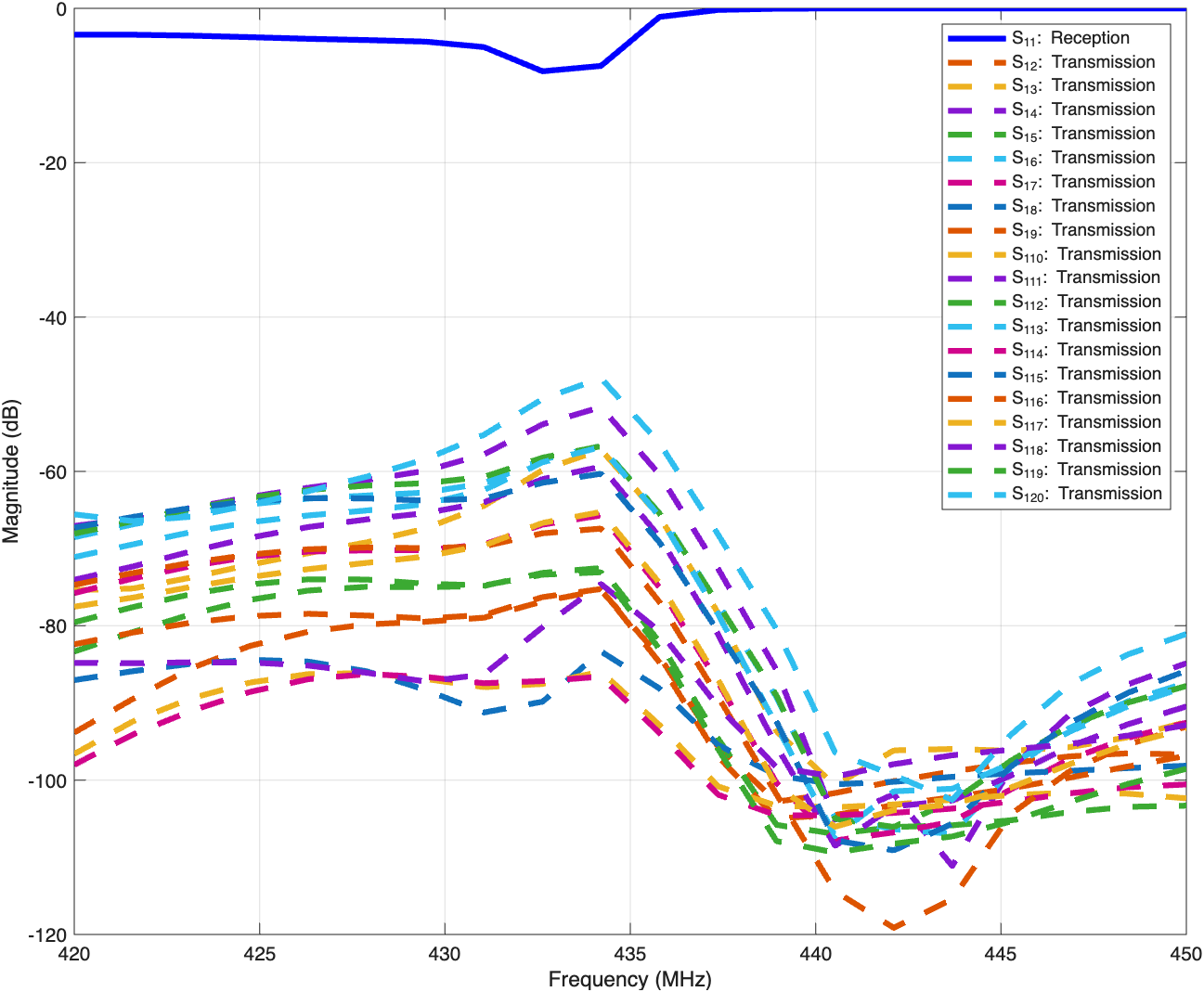}
    \caption{Case 3: Reception and transmission spectra of a 20-element yagi-UHF random array.}
    \label{tranmissionSpectraUHF}
\end{figure}

\begin{figure}[b!]
    \centering
    \includegraphics[width=0.9\columnwidth,clip=true]{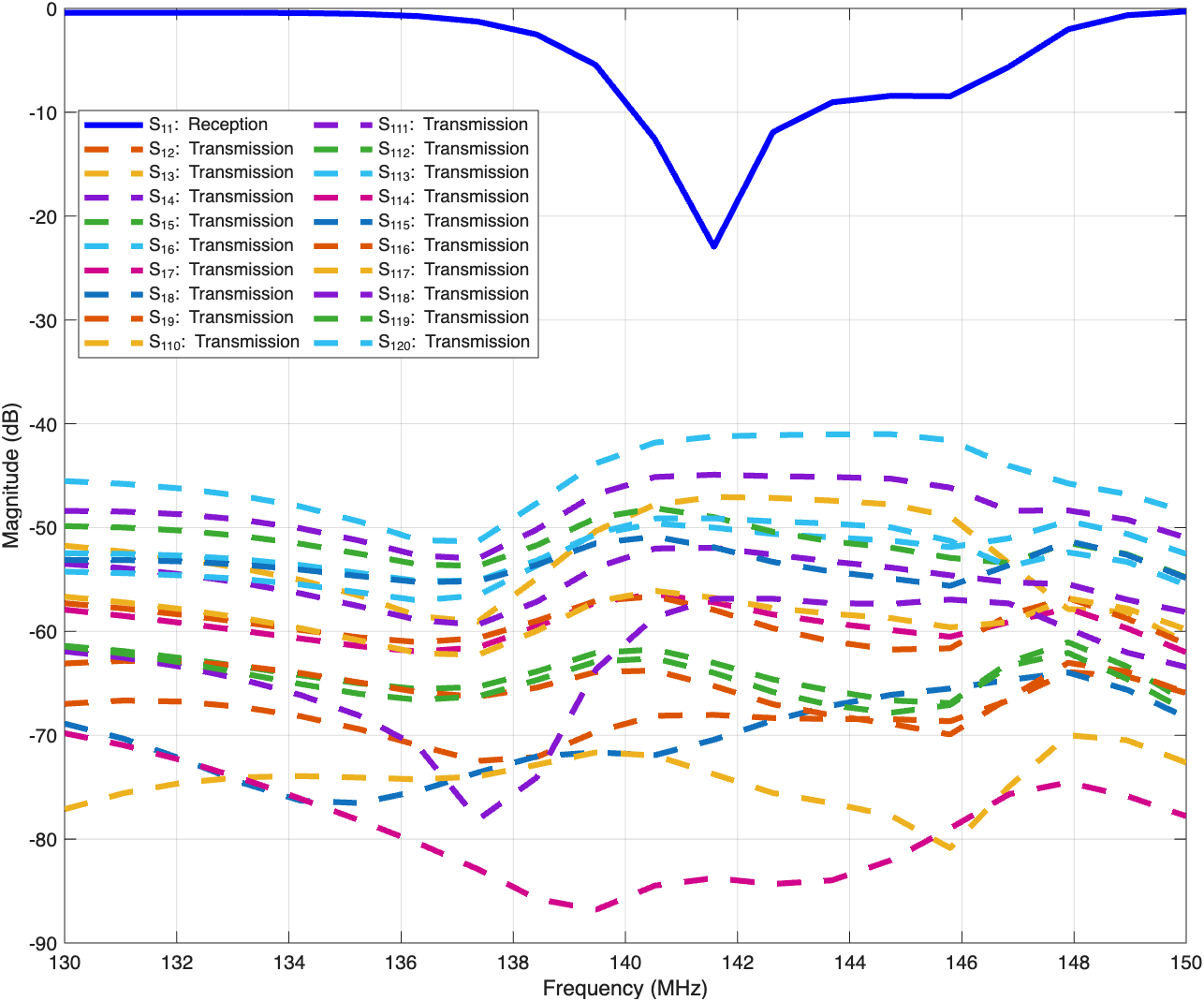}
    \caption{Case 3: Reception and transmission spectra of a 20-element yagi-VHF random array.}
    \label{tranmissionSpectraVHF}
\end{figure}

\subsubsection{Case 3: Transmission spectra on random UHF/VHF array.}
Reception spectra were calculated among all elements of a 20-element random UHF/VHF random array. We will use a 3m separation distance between antennas because the yagi-UHF/VHF antennas are mounted on a 2m cross-boom and that distance is necessary to perform H-plane mechanical steering.

Since we are using a sparse array with typical separations $4.35\lambda_{UHF}$ and $1.45\lambda_{VHF}$ we do not see any differences with the behavior of all antenna's reception spectra. Therefore on the transmission spectra plots, we will include a reception spectra measurement of a single antenna, at the center of the array, to use as a reference. 

Figs.~\ref{tranmissionSpectraUHF} and~\ref{tranmissionSpectraVHF} show the transmission spectra of the referenced antenna against all the remaining antennas; we also show the reception spectrum of the reference antenna to indicate if there is significant interference. We can see that because all of these antennas are separated by 3 meters, which is a large multiple of the wavelength, there is no cross-talk interfering with the reference antenna.



\begin{figure}[t!]
    \centering
    \includegraphics[width=0.9\columnwidth,clip=true]{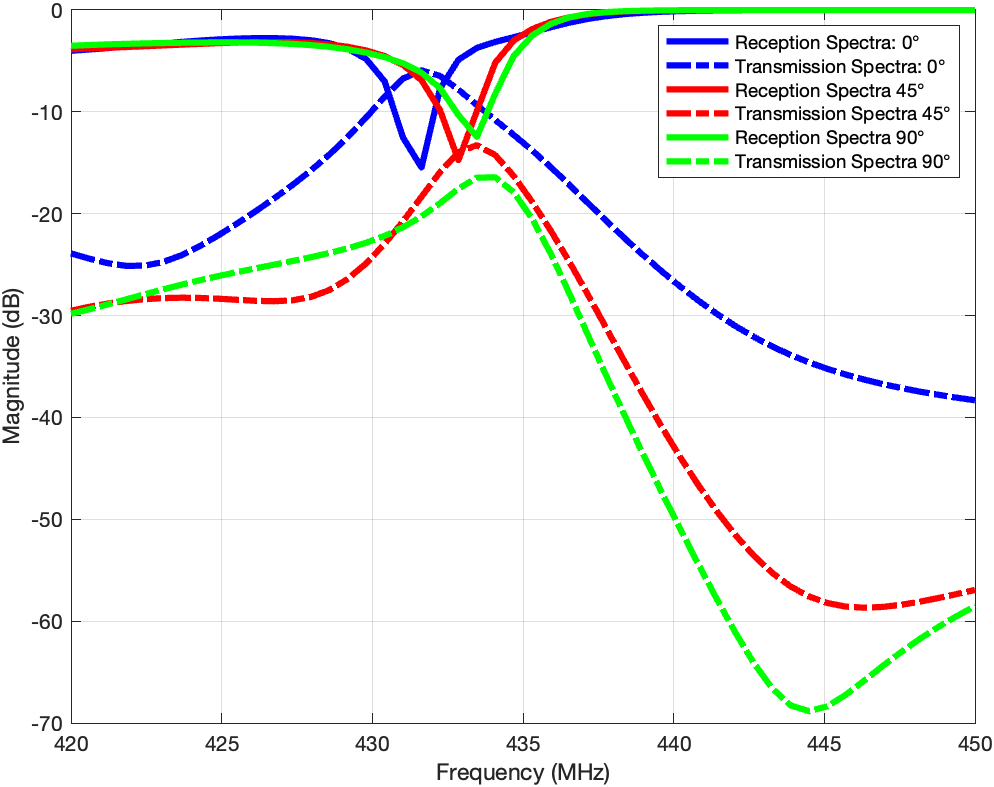}
    \caption{Case 4: Directional axis rotation transmission spectra of a 2-element yagi-UHF antenna array of $0.6\lambda_{UHF}$ separation distance. Fixed $\theta=90\degree$. $\psi$ sweep $(0\degree, 45\degree, 90\degree)$.}
    \label{TspecUHF}
\end{figure}

\begin{figure}[b!]
    \centering
    \includegraphics[width=0.9\columnwidth,clip=true]{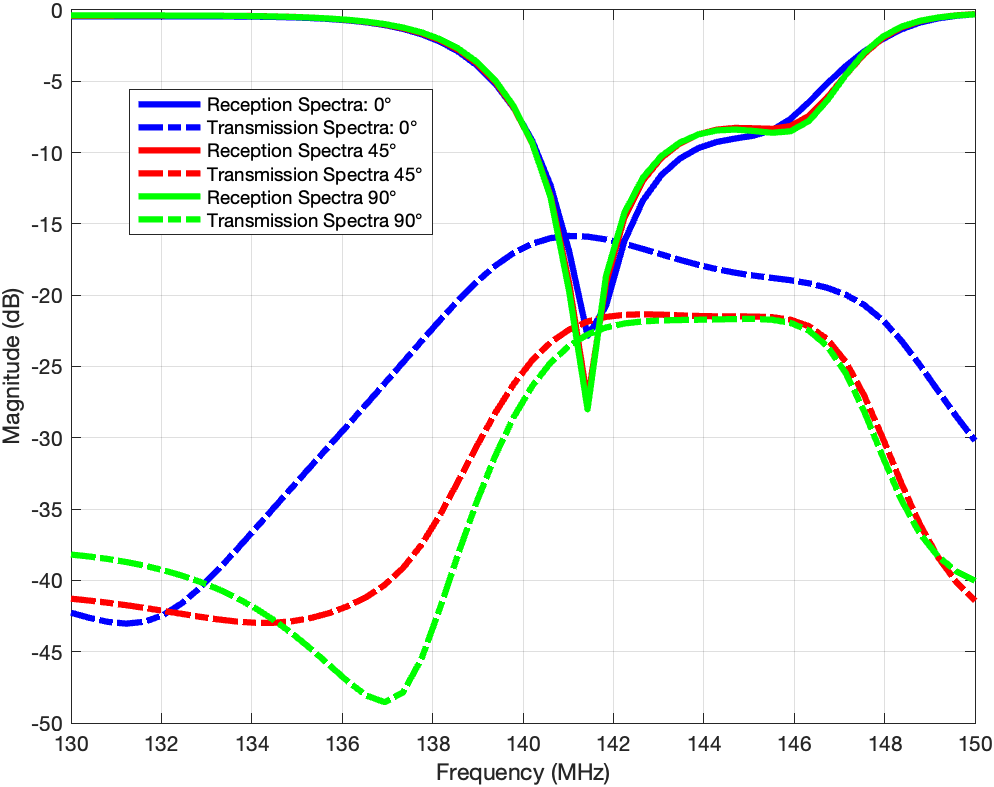}
    \caption{Case 4: Directional axis rotation transmission spectra of a 2-element yagi-VHF antenna array of $0.6\lambda_{VHF}$ separation distance. Fixed $\theta=90\degree$. $\psi$ sweep $(0\degree, 45\degree, 90\degree)$.}
    \label{TspecVHF}
\end{figure}

\begin{figure}[t!]
    \centering
    \includegraphics[width=0.9\columnwidth,clip=true]{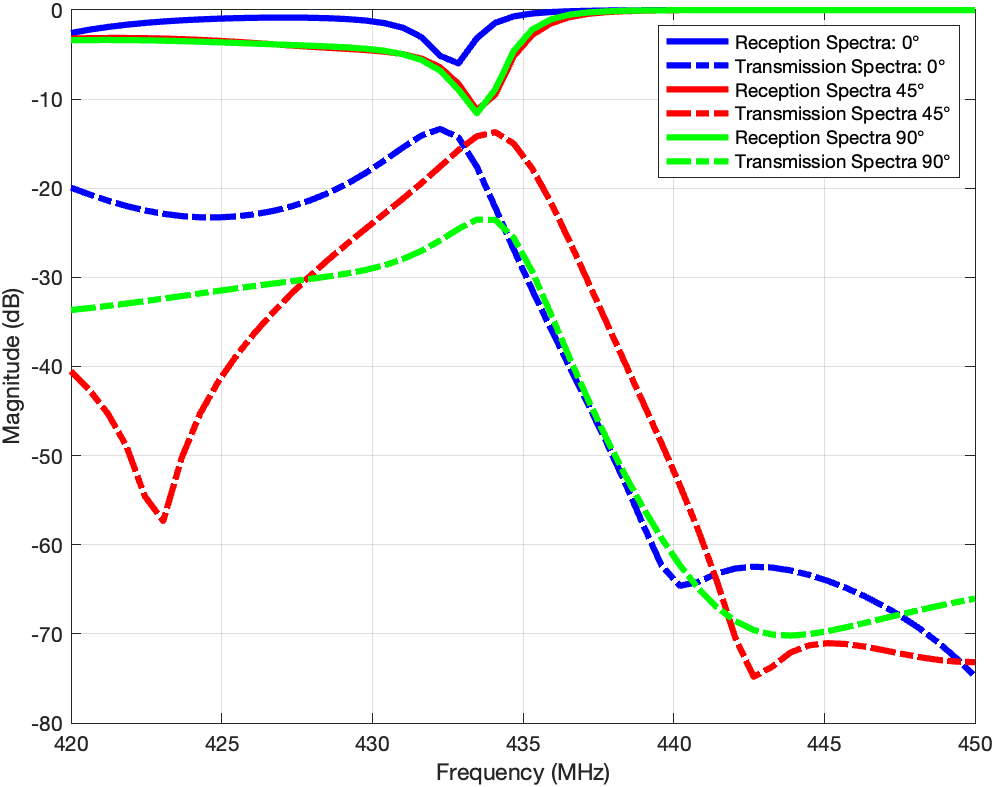}
    \caption{Case 5: Transmission spectra of a 2-element yagi-UHF antenna. $\theta$ sweep $(0\degree, 45\degree, 90\degree)$. Fixed $\phi=0\degree$. Fixed $\psi$.}
    \label{TspecUHFaz0Elsweep}
\end{figure}

\begin{figure}[b!]
    \centering
    \includegraphics[width=0.9\columnwidth,clip=true]{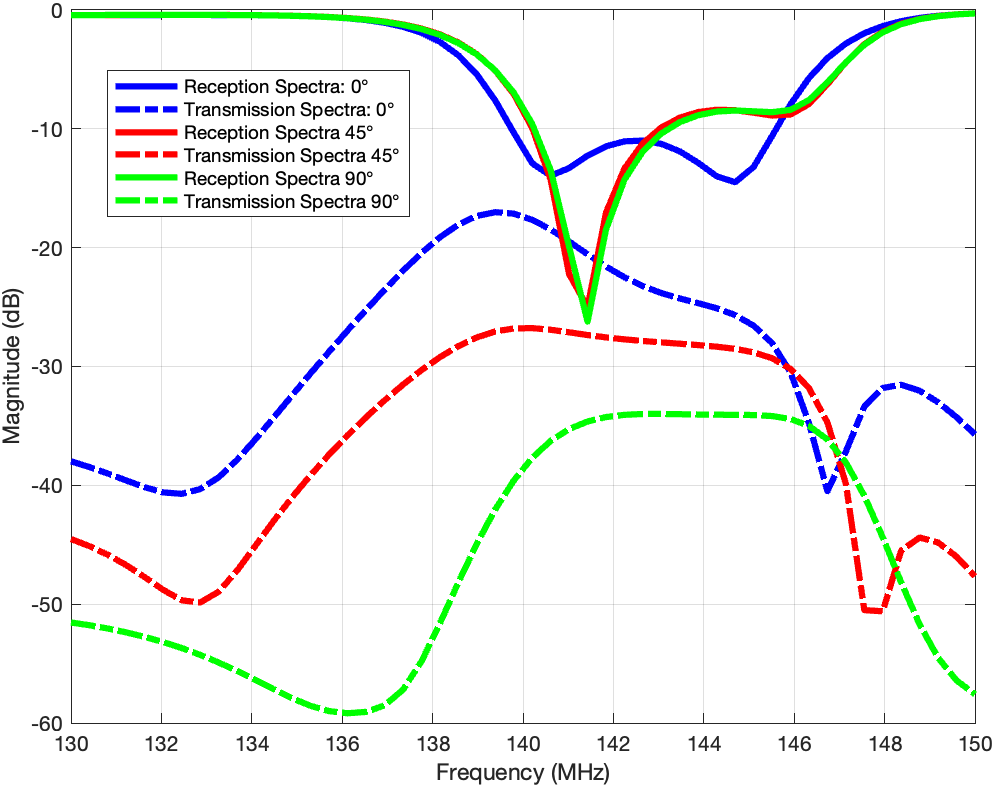}
    \caption{Case 5: Transmission spectra of a 2-element yagi-VHF antenna. $\theta$ sweep $(0\degree, 45\degree, 90\degree)$. Fixed $\phi=0\degree$. Fixed $\psi$.}
    \label{TspecVHFaz0Elsweep}
\end{figure}

\begin{figure}[t!]
    \centering
    \includegraphics[width=0.9\columnwidth,clip=true]{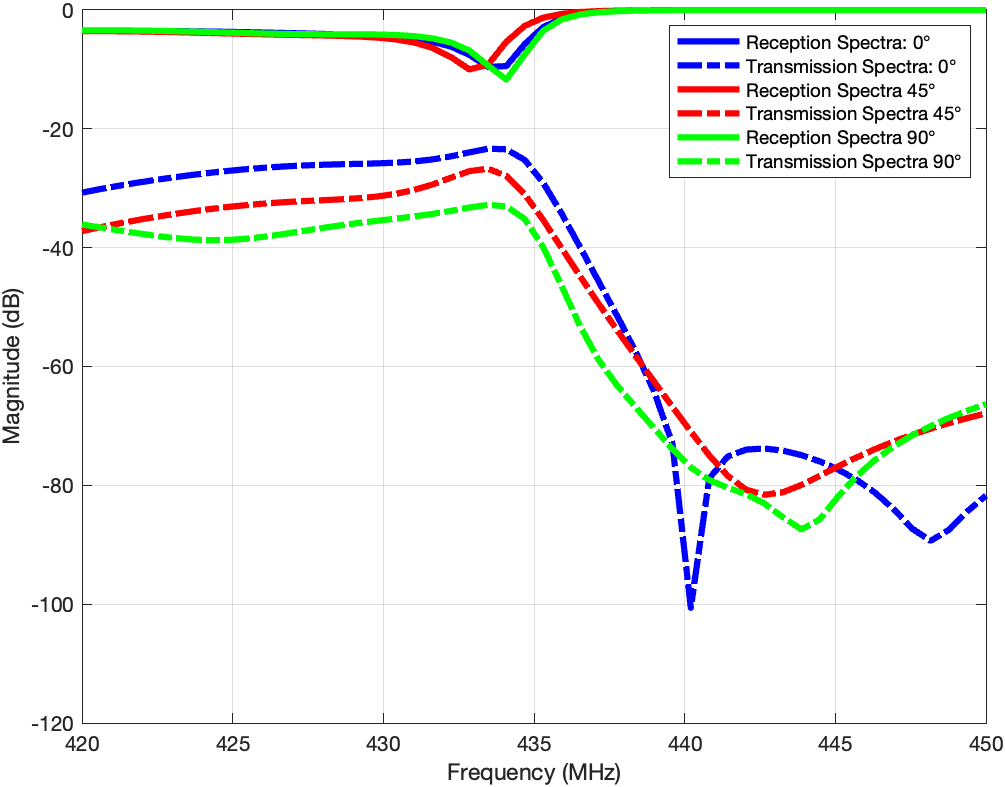}
    \caption{Case 6: Transmission spectra of a 2-element yagi-UHF antenna. Fixed $\theta=60\degree$. $\phi$ sweep $(0\degree, 45\degree, 90\degree)$. Fixed $\psi$.}
    \label{TspecUHFel60phiRot}
\end{figure}

\begin{figure}[b!]
    \centering
    \includegraphics[width=0.9\columnwidth,clip=true]{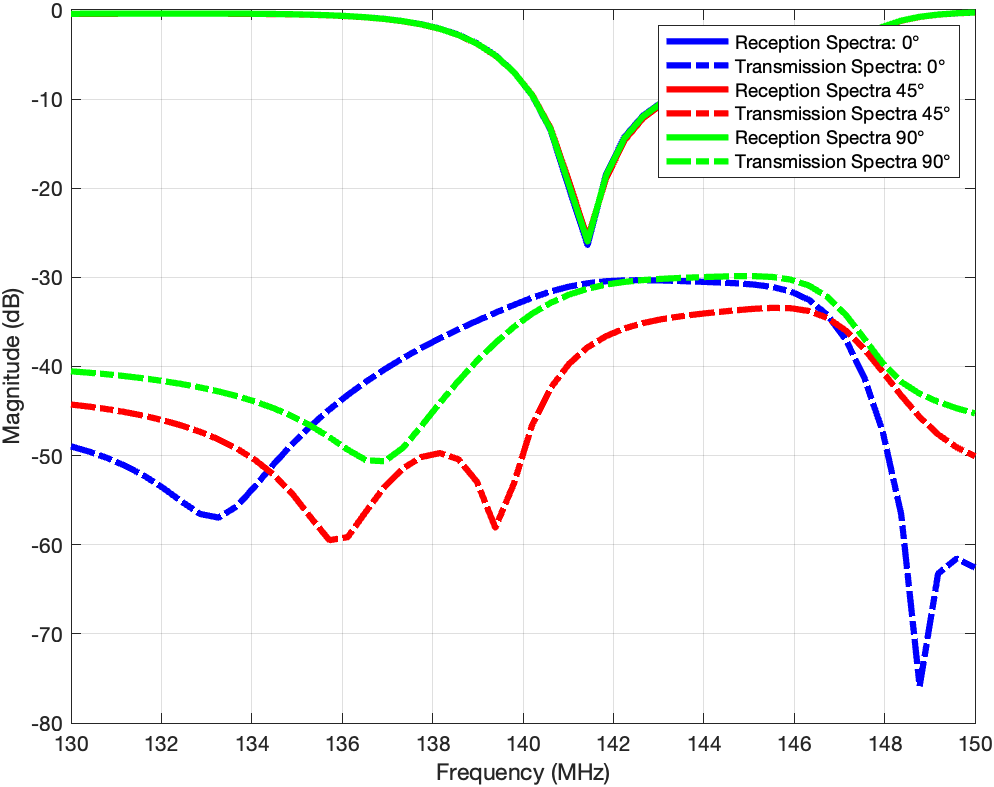}
    \caption{Case 6: Transmission spectra of a 2-element yagi-VHF antenna. Fixed $\theta=60\degree$. $\phi$ sweep $(0\degree, 45\degree, 90\degree)$. Fixed $\psi$.}
    \label{TspecVHFel60phiRot}
\end{figure}

\subsubsection{Case 4: Directional axis rotation}
Fig.~\ref{AxisFig} describes the axis of rotation for the E-plane ($\theta$), H-plane ($\phi$), and directional axis rotation ($\psi$). The antennas will have a directional axis orientation of a $90\degree$ on the E-plane angle and $\psi$ rotations of $0\degree$, $45\degree$, and $90\degree$.

Fig.~\ref{TspecUHF} shows the transmission spectra of our 2-element yagi-UHF array with a separation distance of $0.6\lambda_{UHF}$. We observe that these directional-polarized antennas' rotation affects both the reception and transmission spectra. By rotating $\psi$ from $0\degree$ to $90\degree$ we see a frequency shift towards the right on both the reception and transmission spectra. At $\psi=0\degree$ the antennas are aligned linearly with a small gap and basically it acts as a bigger dipole, therefore having a strong transmission and frequency shift. At $\psi=90\degree$ the antennas are parallel aligned so that coupling gets reduced and its resonant frequency shifts back to its original place.

Fig.~\ref{TspecVHF} depicts the transmission spectra of our 2-element yagi-VHF array with a separation distance of $0.6\lambda_{VHF}$.
At these frequencies there is no significant shift in the reception spectrum by rotating the antennas. In all rotation cases, transmission interference is present on the reception spectra. This means that at this distance $\psi$ rotation is invariant on the resonant frequency.

\subsubsection{Case 5: Directional axis ($\phi=0\degree$) translation without rotation application.}

Fig.~\ref{TspecUHFaz0Elsweep} and Fig.~\ref{TspecVHFaz0Elsweep} will study the transmission spectra effects on a 2-element UHF/VHF simulation where these antennas are pointing at $\theta=[0\degree, 45\degree, 90\degree]$ on E-plane, $\phi=0\degree$ on H-plane, and no $\psi$ directional axis rotation. 

We can observe from both yagi-UHF/VHF that if the antennas are pointing at $\theta=90\degree$ that's the point of most interference, although the transmission is still less than the reception resonance. And the more we tilt down in  the E-plane, the transmission decreases.

\subsubsection{Case 6: Directional axis rotation and translation application.}

Fig.~\ref{TspecUHFel60phiRot} and Fig.~\ref{TspecVHFel60phiRot} show the transmission spectra of a 2-element yagi-UHF/VHF simulation where these antennas pointing at $\theta=60\degree$ on E-plane, $\psi=[0\degree, 45\degree, 90\degree]$ on H-plane, and no $\psi$ directional axis rotation.

For both cases, we can observe that when the antennas are pointing linearly, the transmission spectra increases, but if they point parallel towards the $\phi=90\degree$ then the transmission spectra decreases.

\begin{figure}[t!]
    \centering
    \includegraphics[width=0.9\columnwidth,clip=true]{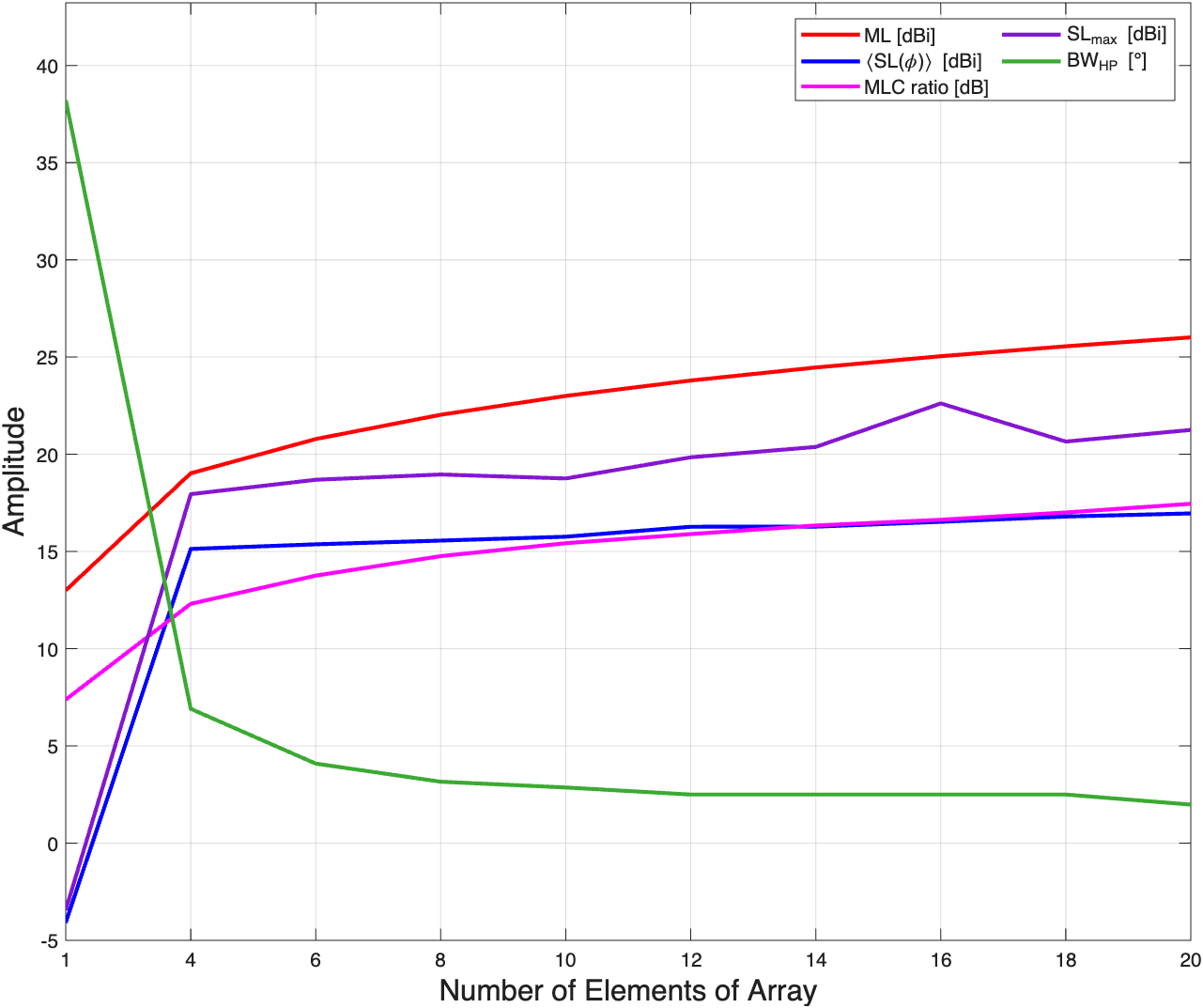}
    \caption{Element sweep analysis for Yagi--UHF antenna random phased array.}
    \label{PaBpIter}
\end{figure}

\begin{figure}[b!]
    \centering
    \includegraphics[width=0.9\columnwidth,clip=true]{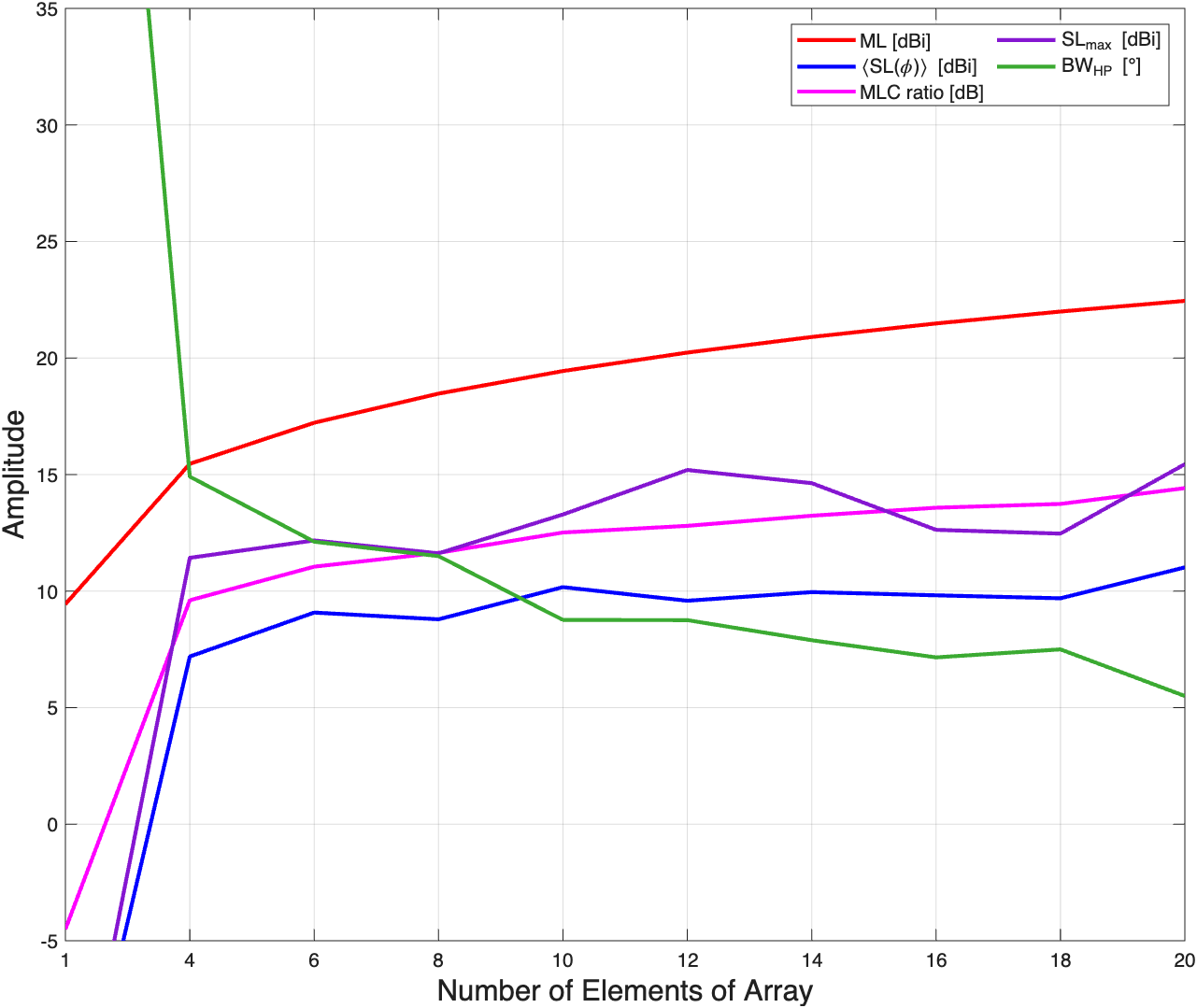}
    \caption{Element sweep analysis for the Yagi--VHF random antenna phased array.}
    \label{PaiterVHF}
\end{figure}

\pagebreak

\section{Radiative Analysis of Yagi-UHF/VHF Antennas}

\subsection{Array Element Sweep Analysis}

Fig.~\ref{UHFbp1v20} compares a single yagi-UHF/VHF antenna with a 20-element pseudo-random phased array. While a single yagi antenna can establish satellite links, it has limitations in data rates and interference rejection due to insufficient sensitivity at elevations below $60\degree$. It also lacks electronic steering capabilities, which would allow better discrimination among closely-spaced satellites. In contrast, the phased array achieves a narrow main lobe (26 dBi for UHF, 22.5 dBi for VHF) and reduced side lobes.

Fig.~\ref{PaBpIter} and Fig.~\ref{PaiterVHF} illustrate how by increasing the number of elements strengthens the main lobe while the side lobe growth rate slows after four elements for the random yagi-UHF/VHF arrays. The $\mathbf{MLC}$ ratio follows a similar increase rate as the main lobe with an attenuation of $\sim10$ dB and $\sim8$ dB, ensuring a $5.4-54.9$x $\mathbf{ML}$-to-$\mathbf{\langle BP \rangle}$ multiplier for the yagi-UHF array and a $5.4-27.5$x $\mathbf{ML}$-to-$\mathbf{\mathbf{\langle BP \rangle}}$ multiplier for the yagi-VHF array. The maximum side lobe stabilizes at 21 dBi for UHF and 16.63 dBi for VHF. Notably, the half-power beam-width shrinks significantly—from 38$\degree$ to 1.5$\degree$ (UHF) and 66.96$\degree$ to 4.5$\degree$ (VHF). This determines the optimal antenna count for minimizing ground signal interference while maintaining strong satellite links, a key factor for multi-beamforming.

\subsection{General Side Lobe Analysis}

This section explores the beam pattern metrics between a uniform and random array. Because the number of elements between arrays is the same, the main lobe for each of the arrangements for the yagi-UHF is 26 dBi, and for the yagi-VHF is 22.5 dBi.

\begin{figure}[t!]
    \centering
    \begin{minipage}[t]{0.4\columnwidth}
        \centering
        \includegraphics[height=0.15\textheight,clip=true]{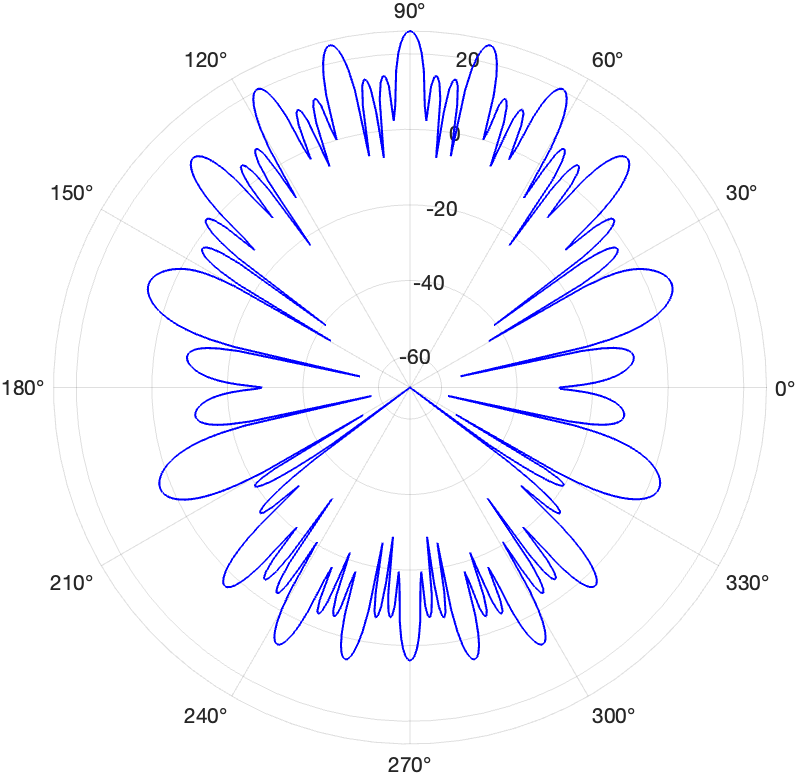}
        \vspace{1ex}
        
        \small Yagi-UHF
    \end{minipage}\hfil
    \begin{minipage}[t]{0.4\columnwidth}
        \centering
        \includegraphics[height=0.15\textheight,clip=true]{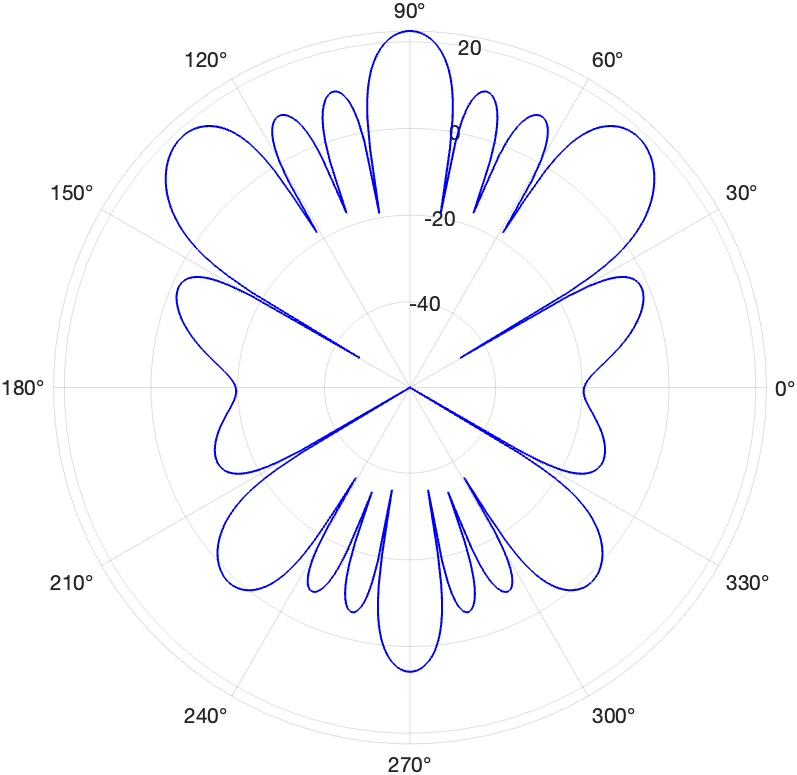}
        \vspace{1ex}
        
        \small Yagi-VHF
    \end{minipage}

    \caption{Uniform yagi-UHF/VHF array beam pattern E-plane cut.}
    \label{uniBPslices}
\end{figure}

\begin{figure}[b!]
    \centering
    \includegraphics[width=0.9\columnwidth,clip=true]{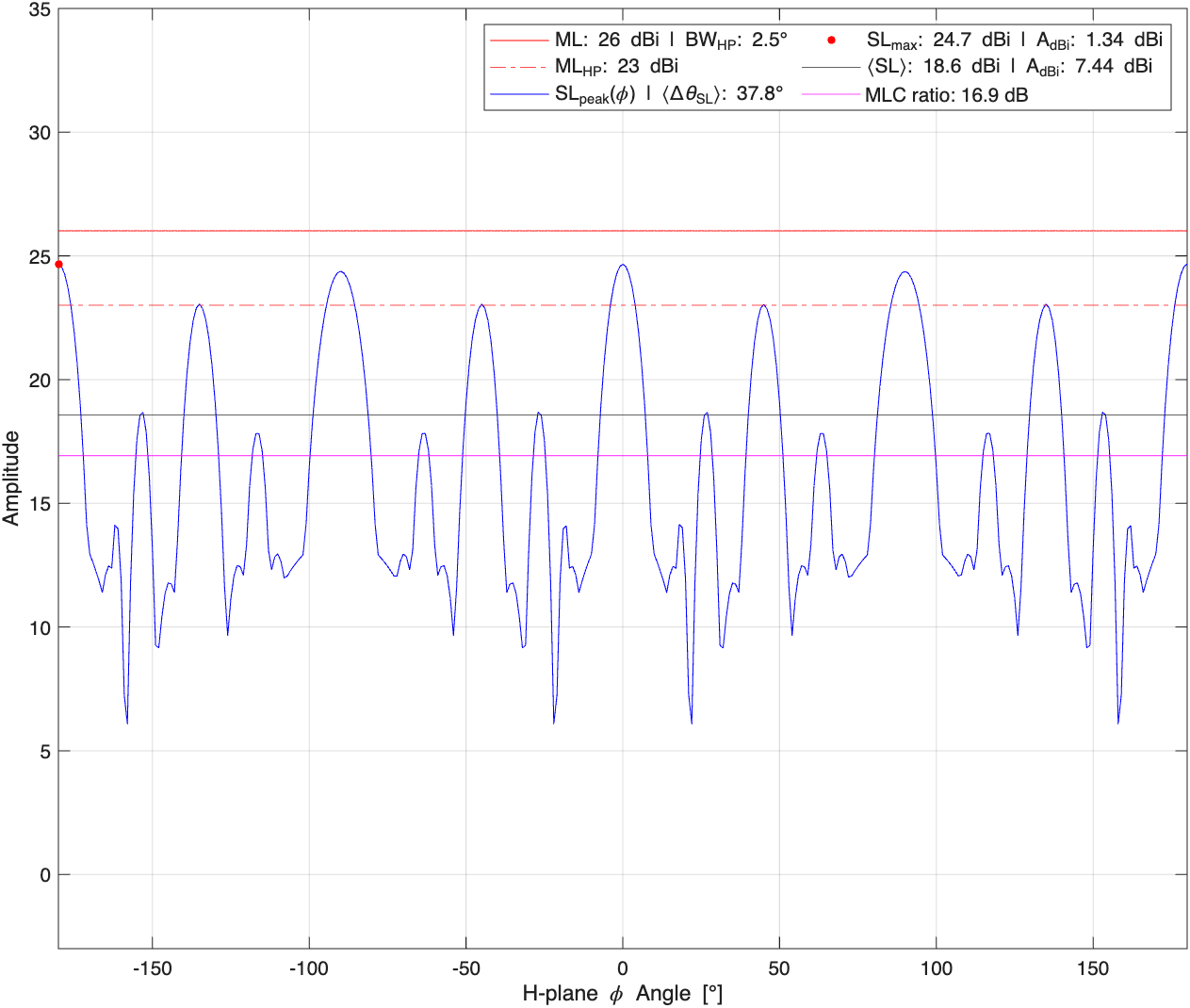}
    \caption{H-plane scan beam pattern metrics for yagi-UHF uniform phased array.}
    \label{uniformSLL}
\end{figure}

\begin{figure}[t!]
    \centering
    \includegraphics[width=0.9\columnwidth,clip=true]{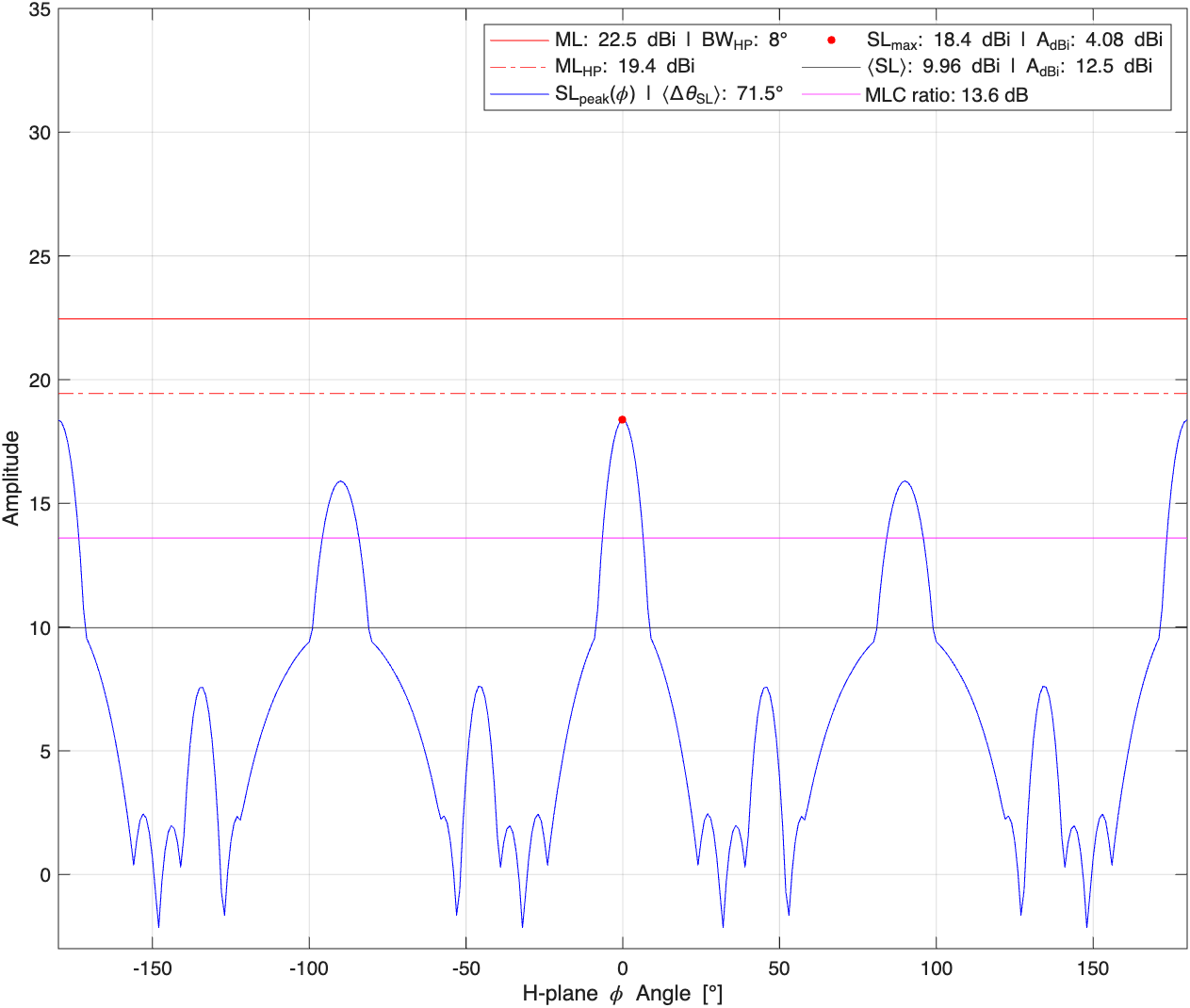}
    \caption{H-plane scan beam pattern metrics for yagi-VHF uniform phased array.}
    \label{uniSLLvhf}
\end{figure}

\subsubsection{Uniform UHF/VHF Array}

\begin{figure}[b!]
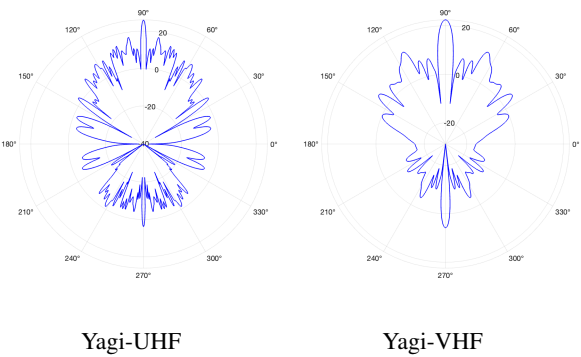

    \centering
    \begin{minipage}[t]{0.4\columnwidth}
        \centering
        \includegraphics[height=0.15\textheight,clip=true]{figures/randBPelUHF.png}
        \vspace{1ex}

        \small Yagi-UHF
    \end{minipage}\hfil
    \begin{minipage}[t]{0.4\columnwidth}
        \centering
        \includegraphics[height=0.15\textheight,clip=true]{figures/randBPelVHF.png}
        \vspace{1ex}

        \small Yagi-VHF
    \end{minipage}

    \caption{Random yagi-UHF/VHF array beam pattern E-plane cuts.}
    \label{randomBPvert}
\end{figure}

Fig.~\ref{uniBPslices} shows the beam pattern E-plane cuts of a yagi-UHF/VHF uniform phased array. At first glance, we observe that these arrangements are not suited for our purposes since there is minimal side lobe attenuation from the main lobe. There are highly sensitive lobes spread out through the elevation plane.

\begin{figure}[t!]
    \centering
    \includegraphics[width=0.9\columnwidth,clip=true]{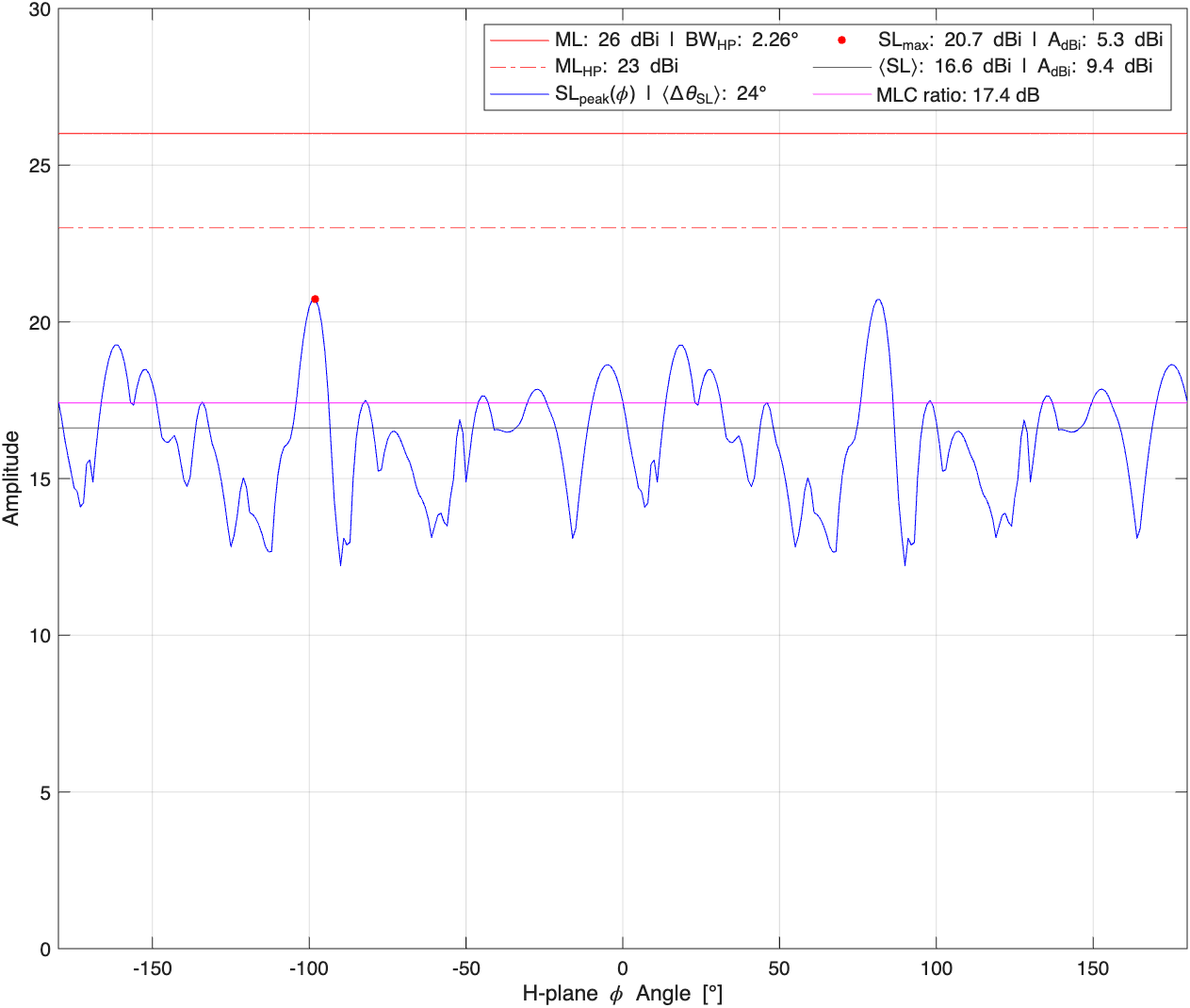}
    \caption{H-plane beam pattern metrics for yagi-UHF random phased array.}
    \label{randomSLL}
\end{figure}

Fig.~\ref{uniformSLL} is the $\mathbf{SL(\phi)}$ for a uniform UHF array. We observe the following results: a $\mathbf{ML}$ of 26 dBi, a $\mathbf{BW_{HP}}$ of 2.5$\degree$, $\mathbf{\Delta \theta_{SL}}$ of 37.8$\degree$, an $\mathbf{SL(\phi_0)_{max}}$ of 24.7 dBi with an attenuation $\mathbf{A_{dBi}}$ of 1.34 dBi, an $\mathbf{\langle SL(\phi) \rangle}$ of 16 dBi with an $\mathbf{A_{dBi}}$ of 9.98 dBi and a $\mathbf{MLC}$ ratio of 16.9 dB.

\begin{figure}[b!]
    \centering
    \includegraphics[width=0.9\columnwidth,clip=true]{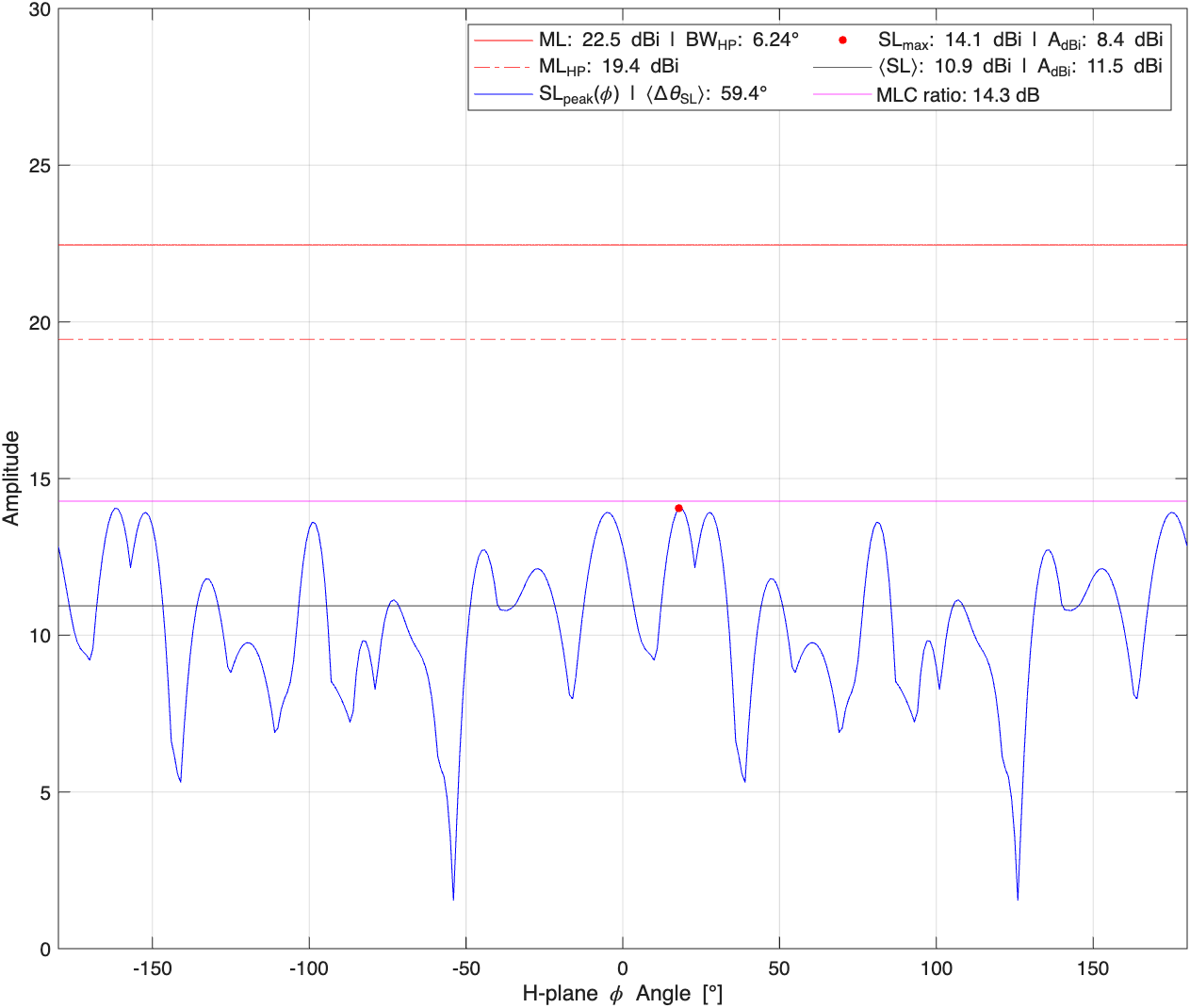}
    \caption{H-plane scan beam pattern metrics for yagi-VHF random phased array.}
    \label{randSLLvhf}
\end{figure}

Fig.~\ref{uniSLLvhf} shows a uniform VHF array's side lobe angular magnitude.  We observe the following results: a $\mathbf{ML}$ of 22.5 dBi, a $\mathbf{BW_{HP}}$ of $8\degree$, $\mathbf{\Delta \theta_{SL}}$ of 71.5$\degree$, an $\mathbf{SL(\phi_0)_{max}}$ of 18.4 dBi with an attenuation $\mathbf{A_{dBi}}$ of 4.08 dBi, an $\mathbf{\langle SL(\phi) \rangle}$ of 6.65 dBi with an $\mathbf{A_{dBi}}$ of 15.8 dBi and a $\mathbf{MLC}$ ratio of 13.6 dB.

Uniform yagi-UHF/VHF arrays are bad for satellite communications, they have low $\mathbf{SL(\phi)}$ attenuation from $\mathbf{ML}$ across the H-plane.

\subsubsection{Random UHF/VHF Array}
The randomness of our array has an inter-element minimum separation distance (3 m) and an array baseline radius (10 m).

Fig.~\ref{randomBPvert} shows the beam pattern E-plane cuts for yagi-UHF/VHF random phased array. At first glance, we observe that these arrangements are best suited for our purposes since we achieve a significant side lobe attenuation from the main lobe.

Fig.~\ref{randomSLL} illustrates the $\mathbf{SL(\phi)}$ performance over the H-plane for a random array of yagi-UHF antennas. We observe the following results: a $\mathbf{ML}$ of 26 dBi, a $\mathbf{BW_{HP}}$ of 2.26$\degree$, $\mathbf{\Delta \theta_{SL}}$ of 24$\degree$, an $\mathbf{SL(\phi_0)_{max}}$ of 20.7 dBi with an attenuation $\mathbf{A_{dBi}}$ of 5.3 dBi, an $\mathbf{\langle SL(\phi) \rangle}$ of 16.6 dBi with an $\mathbf{A_{dBi}}$ of 9.4 dBi and a $\mathbf{MLC}$ ratio of 17.4 dB.

Fig.~\ref{randSLLvhf} illustrates the $\mathbf{SL(\phi)}$ for a random array of yagi-VHF antennas.  We observe the following results: a $\mathbf{ML}$ of 22.5 dBi, a $\mathbf{BW_{HP}}$ of 6.24$\degree$, $\mathbf{\Delta \theta_{SL}}$ of 59.4$\degree$, an $\mathbf{SL(\phi_0)_{max}}$ of 14.1 dBi with an attenuation $\mathbf{A_{dBi}}$ of 8.4 dBi, an $\mathbf{\langle SL(\phi) \rangle}$ of 10.9 dBi with an $\mathbf{A_{dBi}}$ of 11.5 dBi and a $\mathbf{MLC}$ ratio of 14.3 dB.

Random yagi-UHF/VHF arrays are good for satellite communications, they have high $\mathbf{SL(\phi)}$ attenuation from $\mathbf{ML}$ across the H-plane.

\subsection{Beam Steering Analysis}
This section examines steering methods for our phased array, utilizing both electronic and mechanical steering. Individual antenna digitization enables electronic beam steering, while azimuth/elevation rotors allow mechanical adjustments. Simulations will analyze beam steering elevation sweeps, measuring main and side lobes for three configurations: electronic, mechanical, and electronic+mechanical steering. For each, we will assess the E-plane beam steering sweep performance.
\begin{figure}[b!]
    \centering
    \begin{minipage}[t]{0.4\columnwidth}
        \centering
        \includegraphics[height=0.15\textheight,clip=true]{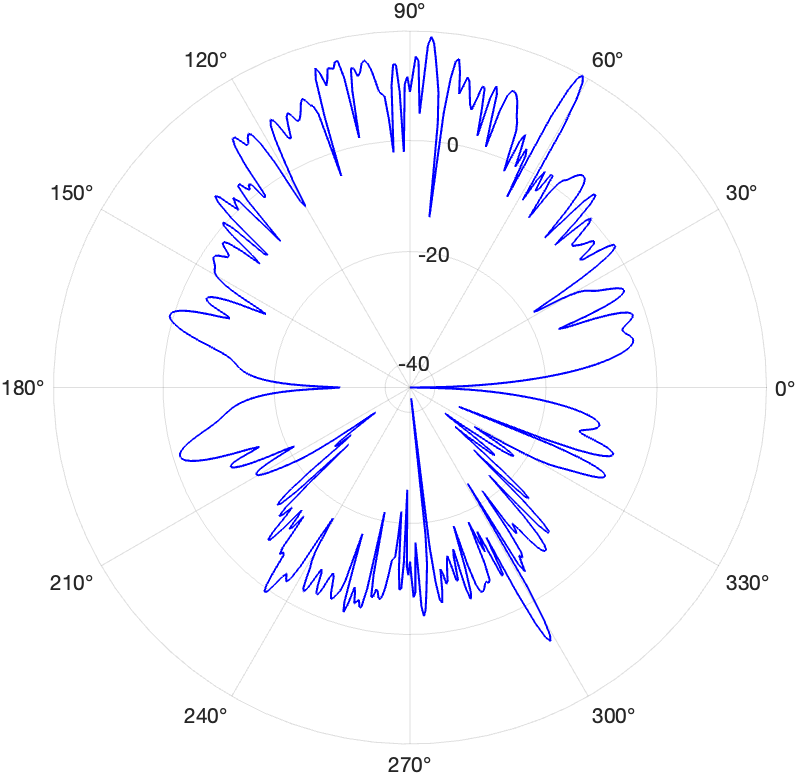}
        \vspace{1ex}

        \small Yagi-UHF
    \end{minipage}\hfil
    \begin{minipage}[t]{0.4\columnwidth}
        \centering
        \includegraphics[height=0.15\textheight,clip=true]{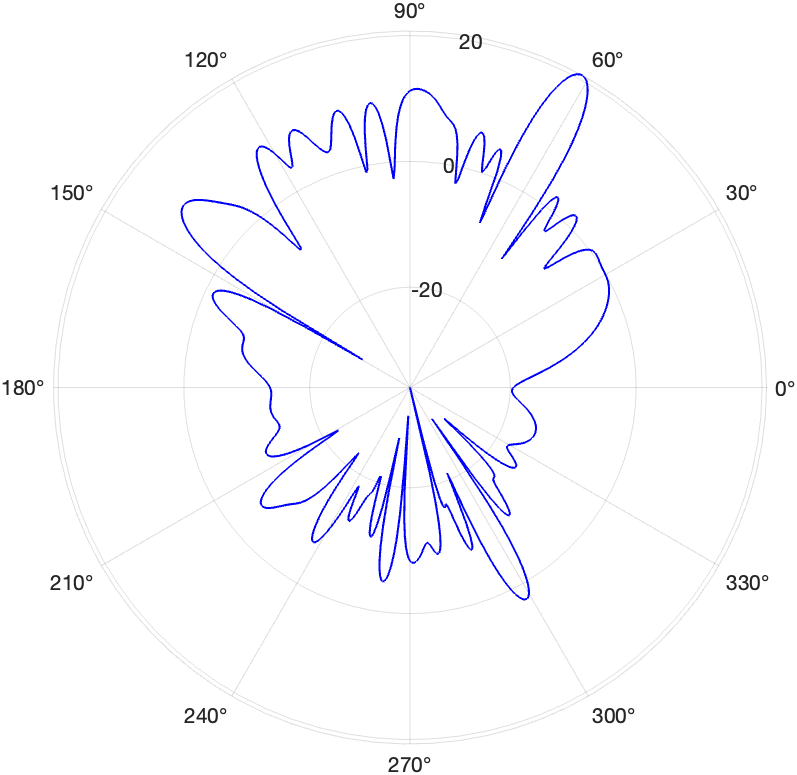}
        \vspace{1ex}

        \small Yagi-VHF
    \end{minipage}

    \caption{E-plane cut for electronic beam steering at 60\degree for random yagi-UHF/VHF phased arrays.}
    \label{Esteer}
\end{figure}

\begin{figure}[t!]
    \centering
    \includegraphics[width=0.9\columnwidth,clip=true]{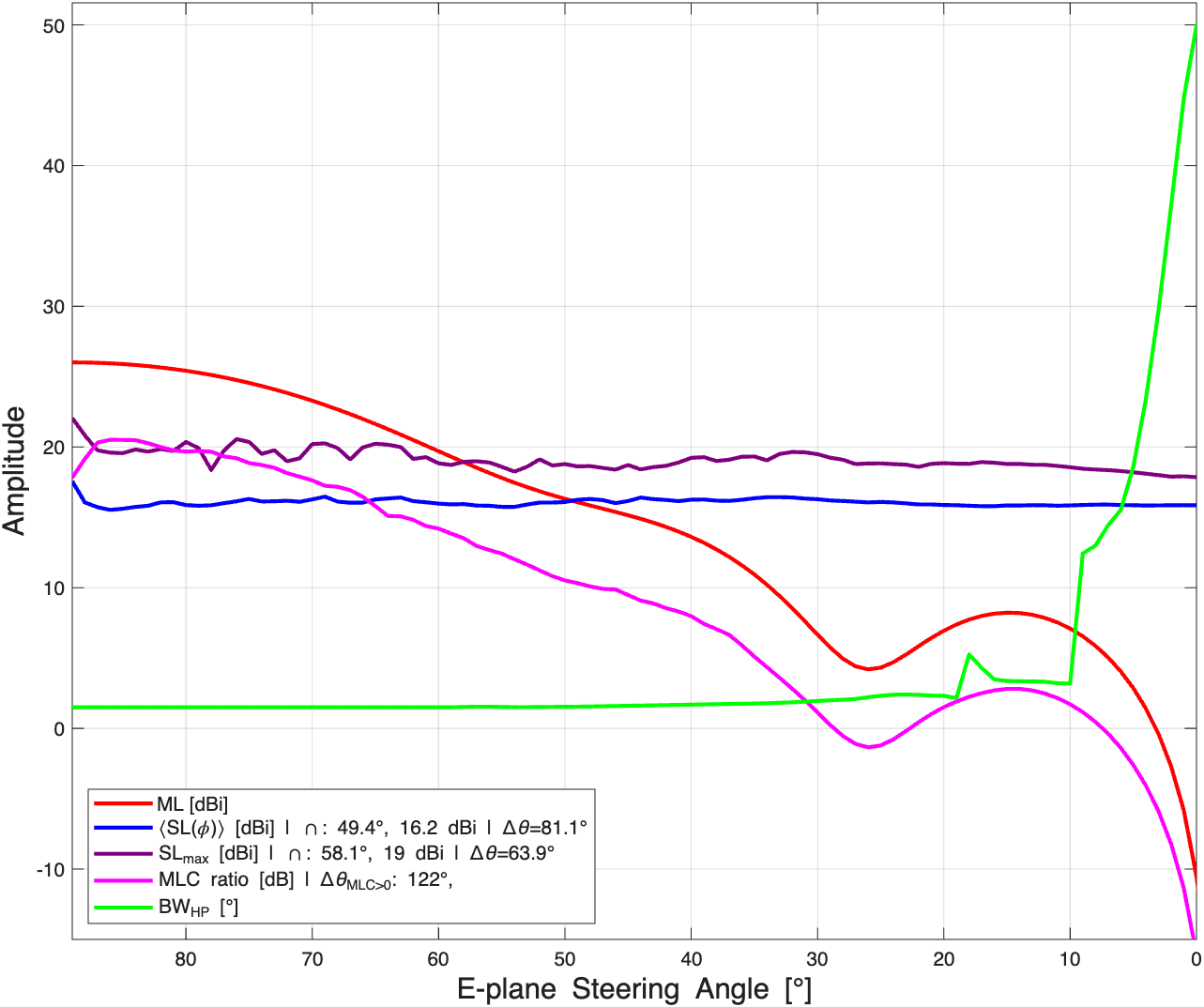}
    \caption{Beam pattern metrics for electronic beam steering sweep for yagi-UHF random phased array.}
    \label{EsteerThetasweep}
\end{figure}

\subsubsection{E-plane Electronic Steering}
This analysis will have the following parameters. Fixed mechanical E-plane $90\degree$ angle. Electronic E-plane steering sweep from $0\degree- 90\degree$. Fixed electronic \& mechanical H-plane steering $0\degree$ angle.

\begin{figure}[b!]
    \centering
    \includegraphics[width=0.9\columnwidth,clip=true]{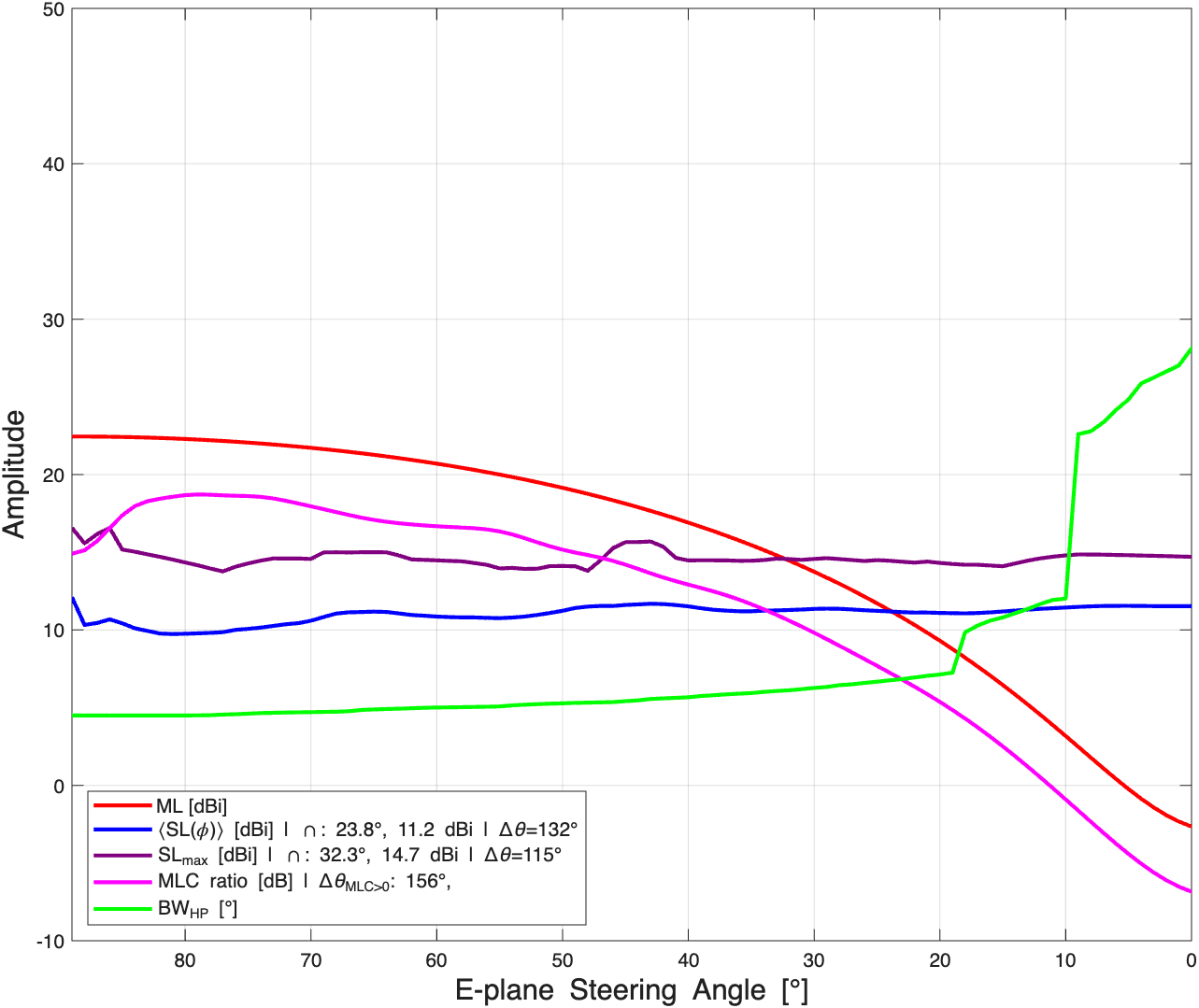}
    \caption{Beam pattern metrics for electronic beam steering sweep for yagi-VHF random phased array.}
    \label{EsteerThetaSweepVHF}
\end{figure}

Fig.~\ref{Esteer} illustrates the beam pattern when electronically steered to 60$\degree$. E-plane beam steering on the phased array's beam pattern has the following effects: All lobes will shift according to the steering angle with the main lobe centered. Lobes pointing at high/low E-plane angles gain/lose intensity.

Fig.~\ref{EsteerThetasweep} shows an E-plane electronic steering sweep evaluating various metrics of array performance for a yagi-UHF random array, with a fixed $\phi=0\degree$ H-plane angle.  We can observe that $\mathbf{ML}$ decreases as the electronic steering angle lowers from the zenith. The $\mathbf{SL(\phi_0)_{max}}$ and $\mathbf{\langle SL(\phi) \rangle}$ remain constant, so the $\mathbf{MLC}$ ratio follows the same behavior as $\mathbf{ML}$. The $\mathbf{BW_{HP}}$ remains roughly constant until it increases sharply for elevations below 10$\degree$ (when it merges with its ``back-lobe'' reflected across the horizontal plane).

We can use these metrics to define an effective ``field of view'' $\Delta\theta$, though this depends on which metric we pick.  
The intersection between $\mathbf{ML}$ and $\mathbf{SL(\phi_0)_{max}}$ occurs at 58.1$\degree$ with a magnitude of 19 dBi, indicating an $\Delta\theta$ E-plane view of 63.9$\degree$ over which the main lobe exceeds \emph{all} sidelobes. The intersection point between $\mathbf{ML}$  and $\mathbf{\langle SL(\phi) \rangle}$ is at 49.4$\degree$ with a magnitude of 16.2 dBi, providing an $\Delta\theta$ E-plane view of 81.1$\degree$ over which the main lobe exceeds the \emph{average} sidelobe. The field of view $\Delta\theta_{MLC>0}$ is $122\degree$.

In Fig.~\ref{EsteerThetaSweepVHF}, we observe similar effects in the behavior of the VHF array. The $\mathbf{ML}$ decreases with elevation, while $\mathbf{SL(\phi_0)_{max}}$ and $\mathbf{\langle SL(\phi) \rangle}$ remain constant, so the $\mathbf{MLC}$ ratio decreases in the same manner as $\mathbf{ML}$. The $\mathbf{BW_{HP}}$ metric increases gradually with an abrupt rise for elevations below 20$\degree$, again due to the (wider) VHF main lobe merging with its back-lobe reflected across the horizontal plane. The intersection between $\mathbf{ML}$ and $\mathbf{SL(\phi_0)_{max}}$ occurs at 32.3$\degree$ with a magnitude of 14.7 dBi, indicating an effective E-plane view range of 115$\degree$. The intersection point between $\mathbf{ML}$  and $\mathbf{\langle SL(\phi) \rangle}$ is at 23.8$\degree$ with a magnitude of 11.2 dBi, providing an $\Delta\theta$ E-plane view of 132$\degree$. The field of view $\Delta\theta_{MLC>0}$ is $156\degree$.

\begin{figure}[t!]
    \centering
    \includegraphics[width=0.9\columnwidth]{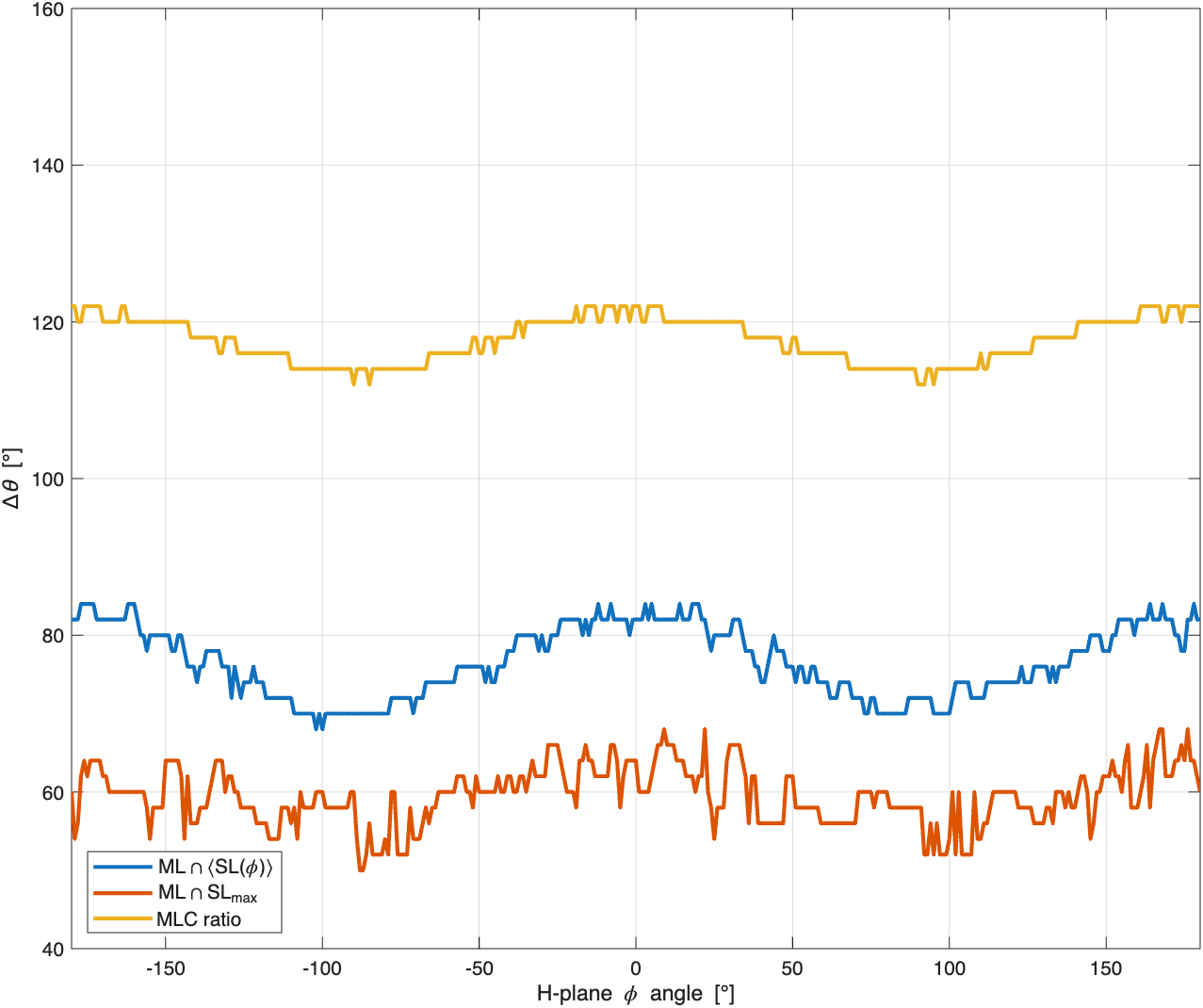}
    \vspace{1ex}

    \caption{$\Delta\theta$ for $\mathbf{\langle SL(\phi)\rangle}$ and $\mathbf{SL(\phi_0)_{max}}$ for a yagi-UHF random phased array | E-plane beam steering sweeps through H-plane.}
    \label{AzDthetaUHF}
\end{figure}

\begin{figure}[b!]
    \centering
    \includegraphics[width=0.9\columnwidth]{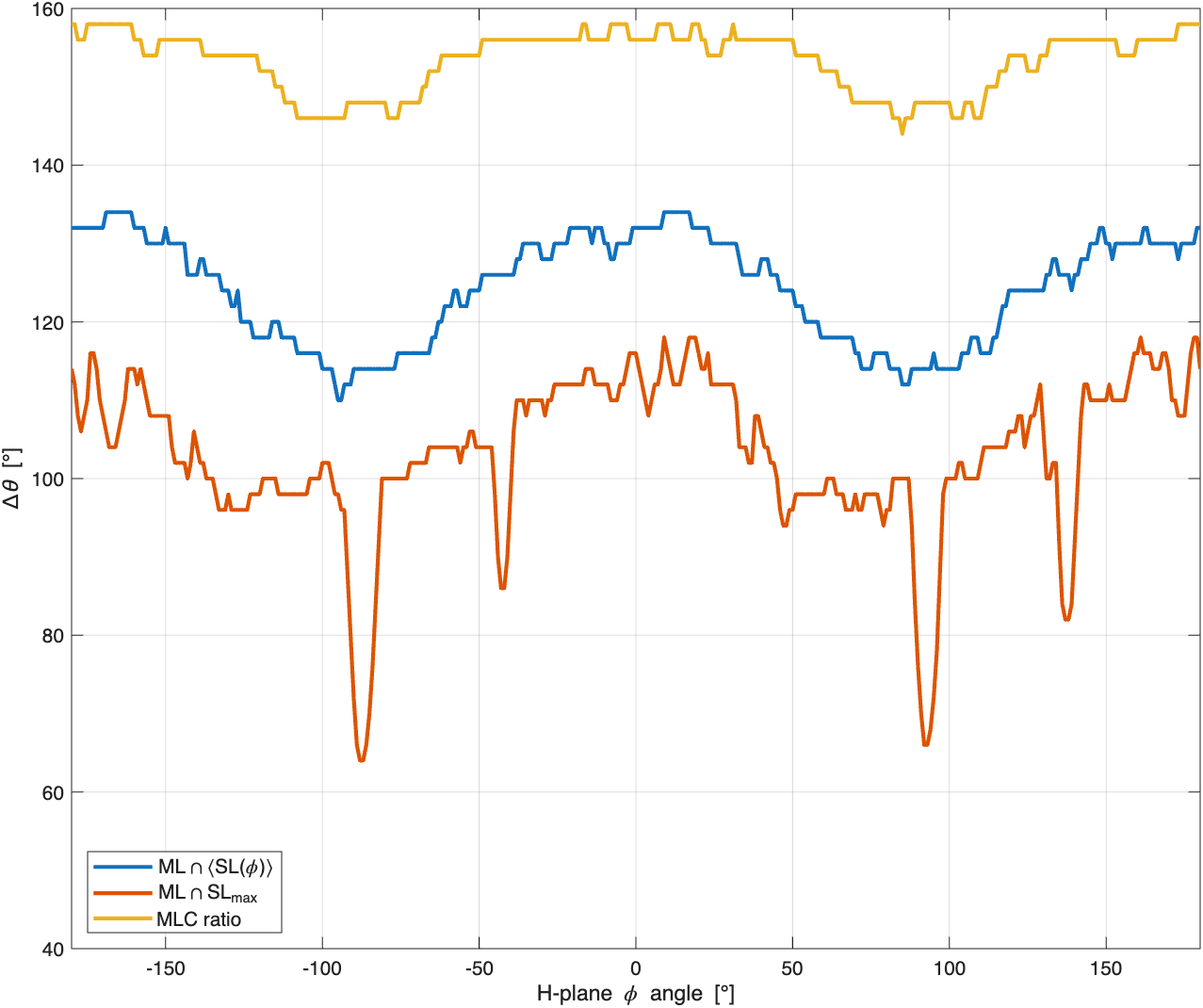}
    \vspace{1ex}

    \caption{$\Delta\theta$ for $\mathbf{\langle SL(\phi)\rangle}$ and $\mathbf{SL(\phi_0)_{max}}$ for a yagi-VHF random phased array | E-plane beam steering sweeps through H-plane.}
    \label{AzDthetaVHF}
\end{figure}

\begin{figure}[t!]
    \centering
    \begin{minipage}[t]{0.45\columnwidth}
        \centering
        \includegraphics[width=\columnwidth,clip=true]{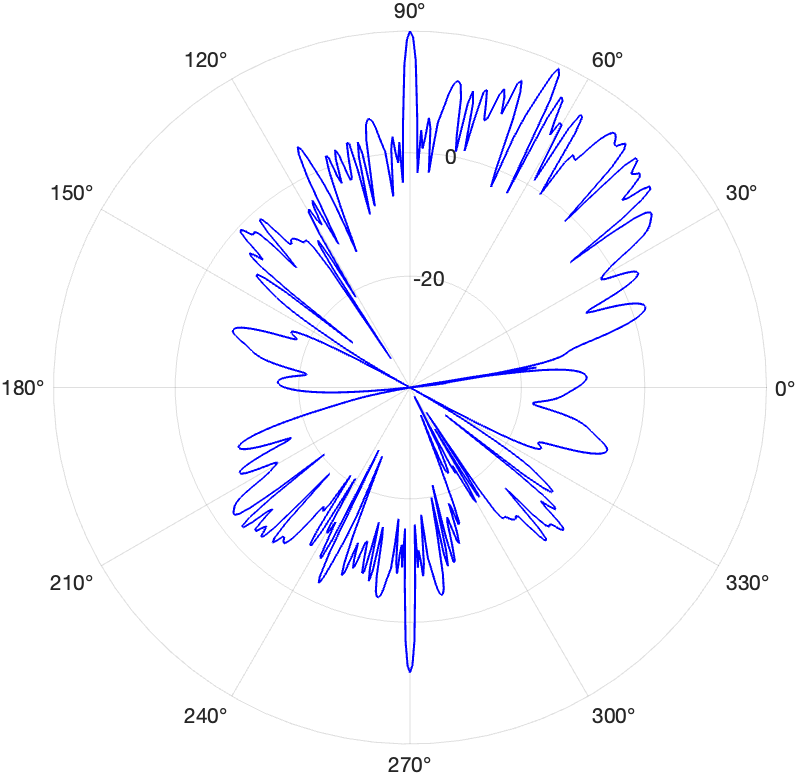}
        \vspace{1ex}

        \small Yagi-UHF
    \end{minipage}
    \hfill
    \begin{minipage}[t]{0.45\columnwidth}
        \centering
        \includegraphics[width=\columnwidth,clip=true]{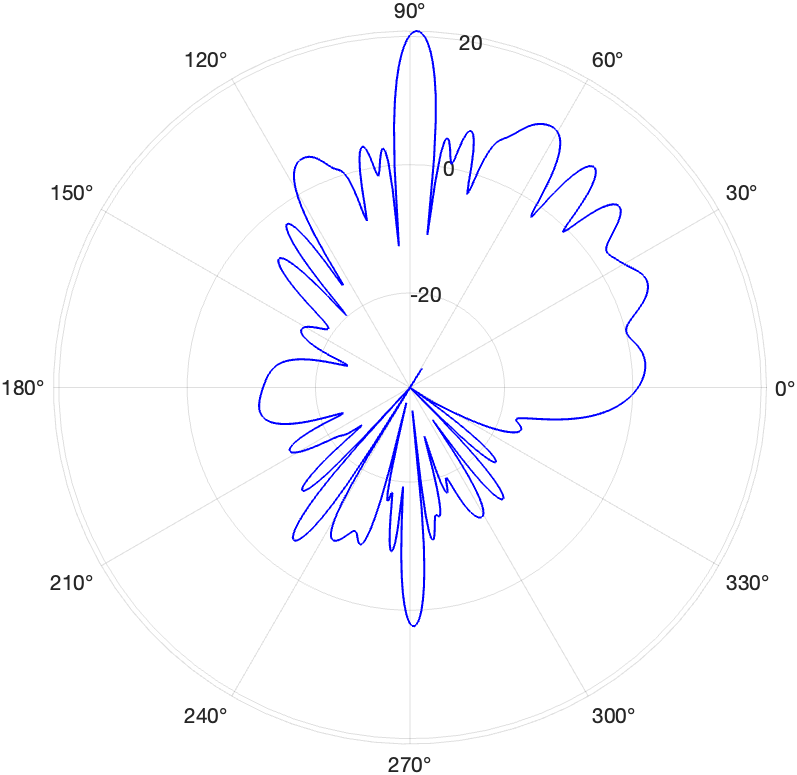}
        \vspace{1ex}

        \small Yagi-VHF
    \end{minipage}

    \caption{E-plane cut for mechanical steering at 60\degree for random yagi-UHF/VHF phased arrays.}
    \label{Msteer}
\end{figure}

Fig.~\ref{AzDthetaUHF} shows how our various ``field of view'' $\Delta\theta$ metrics change for different H-plane slices, for the yagi-UHF random phased array. Ideally, we would like the $\Delta\theta$ metric to have a high value and remain constant across the H-plane, but this is only approximately the case.
We observe slight fluctuations for the random Yagi-UHF scenario. A $\mathbf{\Delta\theta_{\langle SL(\phi)\rangle}}$ of $16\degree$, a $\mathbf{\Delta\theta_{SL_{max}}}$ of $18\degree$, and a $\mathbf{\Delta\theta_{MLC}}$ of $10\degree$. We also have a mean $\Delta\theta_{\langle SL(\phi)\rangle}$ of $76.64\degree$, a mean $\Delta\theta_{SL_{max}}$ of $59.44\degree$, and a mean $\mathbf{\Delta\theta_{MLC}}$ of $117.7\degree$. These variations are small, and we retain a moderate effective field of view across the H-plane.  

Fig.~\ref{AzDthetaVHF} shows similar results for the metric $\Delta\theta$ for $\mathbf{\langle SL(\phi)\rangle}$ and $\mathbf{SL(\phi_0)_{max}}$ for Yagi-VHF random phased arrays for different azimuthal slices.  As with the UHF case, the ideal condition would be a large and constant $\Delta\theta$ across the H-plane.  However, we observe slight higher for the random Yagi-VHF scenario. A $\mathbf{\Delta\theta_{\langle SL(\phi)\rangle}}$ of $24\degree$, a $\mathbf{\Delta\theta_{SL_{max}}}$ of $54\degree$, and a $\mathbf{\Delta\theta_{MLC}}$ of $14\degree$. We also have a mean $\Delta\theta_{\langle SL(\phi)\rangle}$ of $124.21\degree$, a mean $\Delta\theta_{SL_{max}}$ of $103.21\degree$, and a mean $\mathbf{\Delta\theta_{MLC}}$ of $153.29\degree$. Overall the effective field of view is still large, but much more sensitive to H-plane angle.

\subsubsection{Mechanical Steering}

\begin{figure}[b!]
    \centering
    \includegraphics[width=0.9\columnwidth,clip=true]{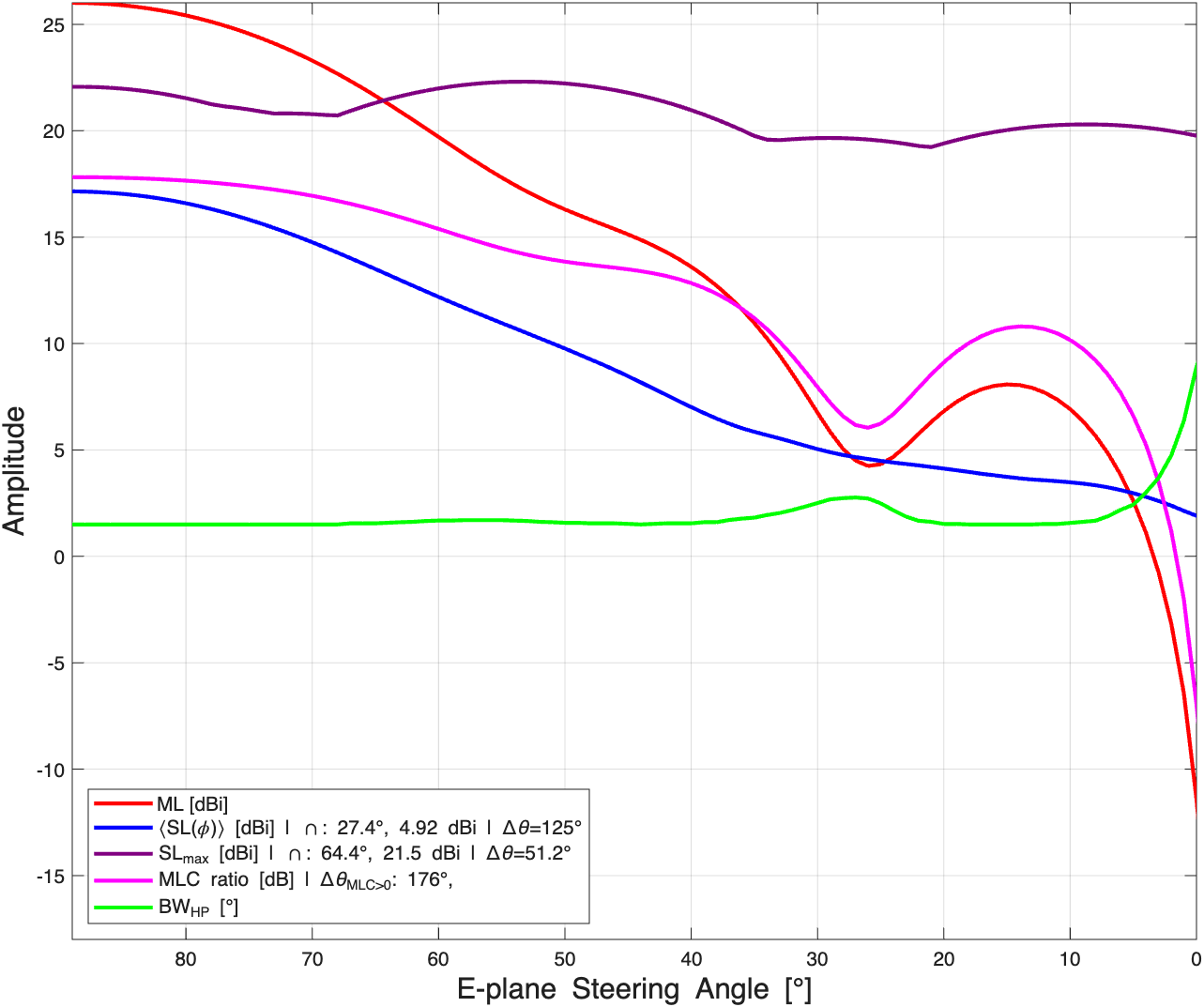}
    \caption{Beam pattern metrics for mechanical steering sweep for yagi-UHF random phased array.}
    \label{MsteerThetasweep}
\end{figure}

\begin{figure}[t!]
    \centering
    \includegraphics[width=0.9\columnwidth,clip=true]{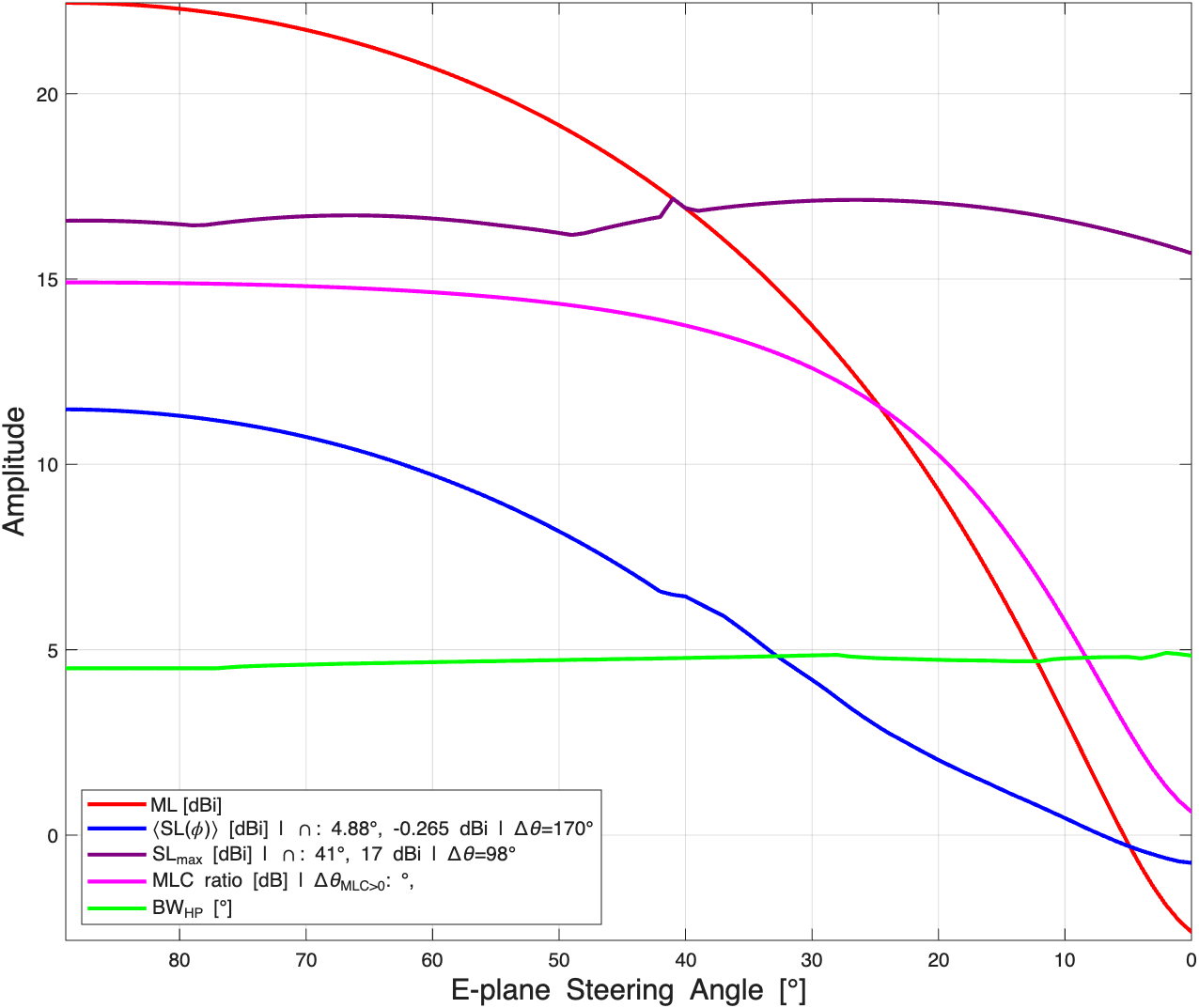}
    \caption{Beam pattern metrics for mechanical steering sweep for yagi-VHF random phased array.}
    \label{MsteerThetaSweepVHF}
\end{figure}

This analysis will have the following parameters. Fixed electronic E-plane $90\degree$ angle. Mechanical E-plane steering sweep from $0\degree- 90\degree$. Fixed electronic \& mechanical H-plane steering $0\degree$ angle.

Fig.~\ref{Msteer} shows the beam pattern mechanically steered to 60$\degree$.  Unlike electronic steering, this method does not affect the relative positioning of lobes within the radiation pattern. However, it introduces different effects: The lobes do not shift as they do in electronic steering. The cosine factor of the individual antenna responses follow the steering angle, while the main lobe remains pointed upward, resulting in attenuation of the main lobe.

In Fig.~\ref{MsteerThetasweep} shows the $\mathbf{ML}$ decreases as the mechanical steering angle lowers from the zenith. The $\mathbf{SL(\phi_0)_{max}}$, and $\mathbf{\langle SL(\phi) \rangle}$ decrease as well, and the $\mathbf{MLC}$ ratio follows the same behavior as $\mathbf{ML}$. The $\mathbf{BW_{HP}}$ remains roughly constant before increasing below 10$\degree$: this increase is more smooth than the case of electronically-steered arrays, as it is not due to the merging of the main lobe and back lobe, but the main lobe spreading into the null of the antenna pattern. The intersection between $\mathbf{ML}$ and $\mathbf{SL(\phi_0)_{max}}$ occurs at 64.4$\degree$ with a magnitude of 21.5 dBi, indicating an $\Delta\theta$ E-plane view of 51.2$\degree$. The intersection point between $\mathbf{ML}$  and $\mathbf{\langle SL(\phi) \rangle}$ is at 27.5$\degree$ with a magnitude of 4.92 dBi, providing an $\Delta\theta$ E-plane view of 125$\degree$. The field of view $\Delta\theta_{MLC>0}$ is $122\degree$.

In Fig.~\ref{MsteerThetaSweepVHF}, we observe similar effects for the VHF array. The $\mathbf{ML}$ decreases, the $\mathbf{SL(\phi_0)_{max}}$ and $\mathbf{\langle SL(\phi) \rangle}$ remains constant, the $\mathbf{MLC}$ ratio has a similar behavior to $\mathbf{ML}$, and the $\mathbf{BW_{HP}}$ metric remains constant. The intersection between $\mathbf{ML}$ and $\mathbf{SL(\phi_0)_{max}}$ occurs at 41$\degree$ with a magnitude of 17 dBi, indicating an effective E-plane view range of 98$\degree$. The intersection point between $\mathbf{ML}$  and $\mathbf{\langle SL(\phi) \rangle}$ is at 4.88$\degree$ with a magnitude of -0.2 dBi, providing an $\Delta\theta$ E-plane view of 170$\degree$. The field of view $\Delta\theta_{MLC>0}$ is $180\degree$.

\subsubsection{Electronic and Mechanical Steering}

\begin{figure}[b!]
    \centering
    \begin{minipage}[t]{0.45\columnwidth}
        \centering
        \includegraphics[width=\columnwidth,clip=true]{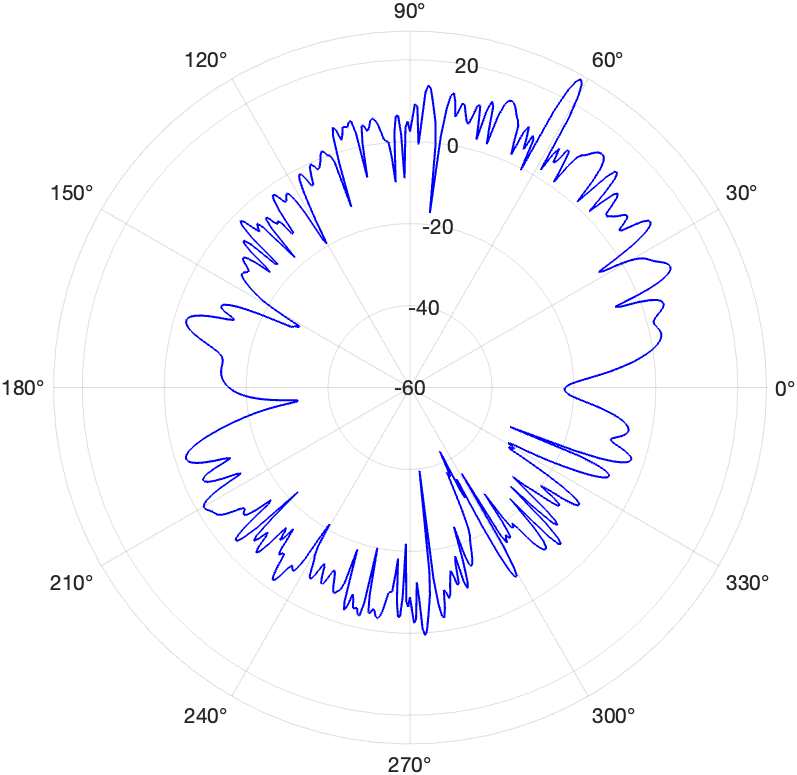}
        \vspace{1ex}

        \small Yagi-UHF
    \end{minipage}
    \hfill
    \begin{minipage}[t]{0.45\columnwidth}
        \centering
        \includegraphics[width=\columnwidth,clip=true]{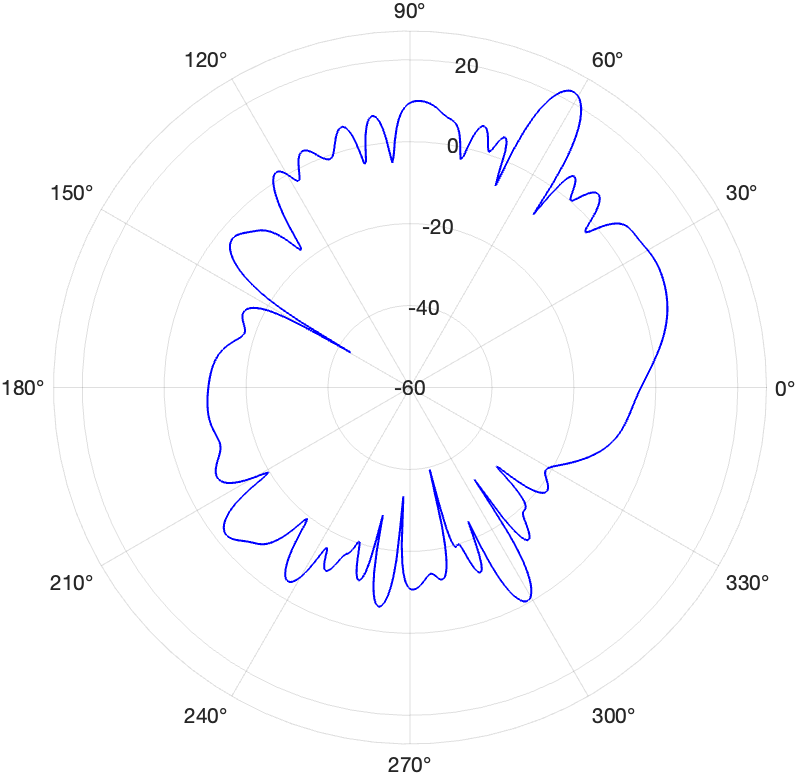}
        \vspace{1ex}

        \small Yagi-VHF
    \end{minipage}

    \caption{E-plane cut for electronic+mechanical steering at 60\degree for random yagi-UHF/VHF phased arrays.}
    \label{EMsteer}
\end{figure}

\begin{figure}[t!]
    \centering
    \includegraphics[width=0.9\columnwidth,clip=true]{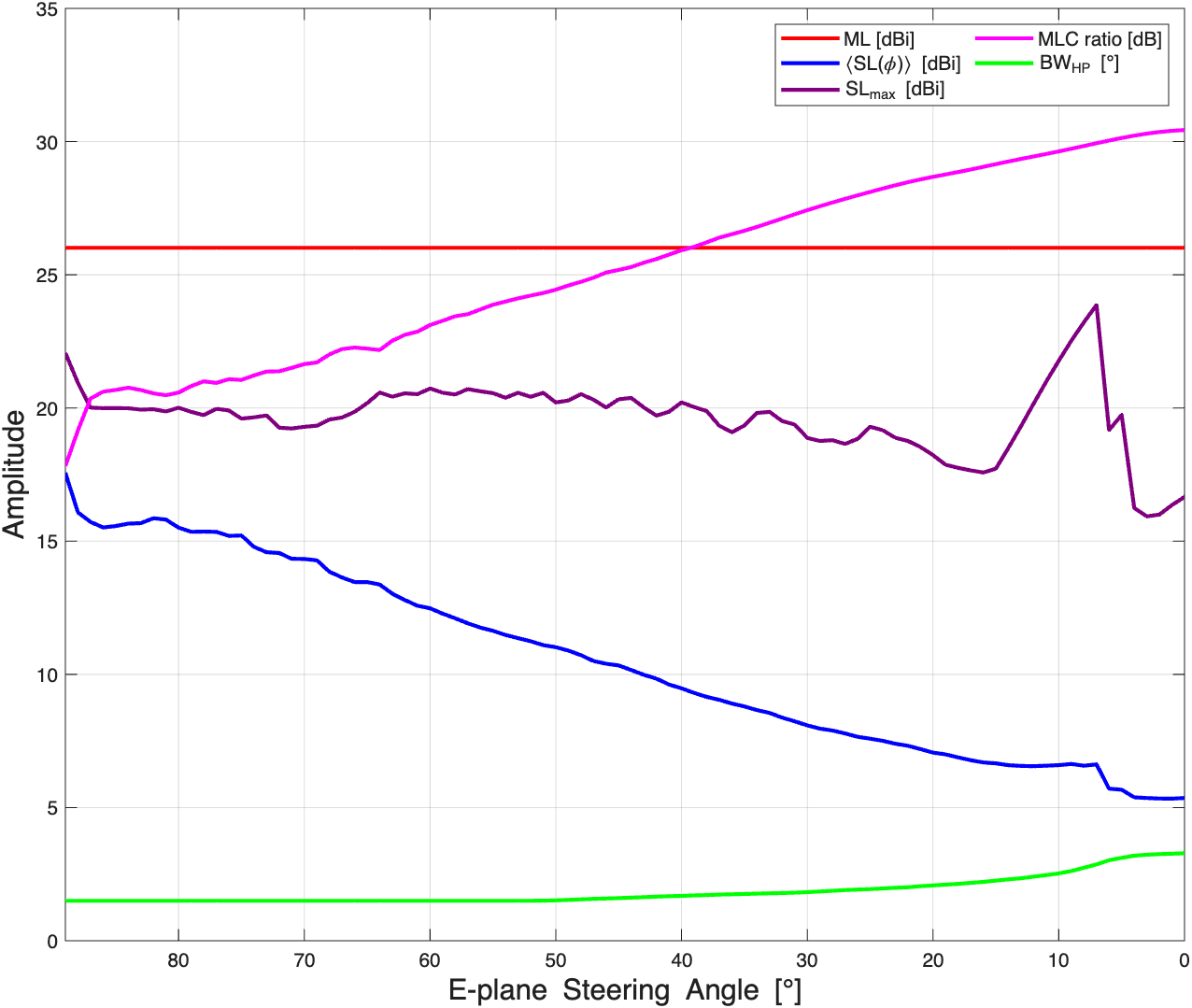}
    \caption{Beam pattern metrics for electronic+mechanical steering sweep for yagi-UHF random phased array.}
    \label{EMsteerThetasweep}
\end{figure}

This analysis will have the following parameters. Electronic \& mechanical E-plane steering sweep from $0\degree- 90\degree$. Fixed  electronic \& mechanical H-plane steering $0\degree$ angle.

Fig.~\ref{EMsteer} illustrates the beam pattern electronically and mechanically steered to 60$\degree$. By integrating both electronic and mechanical steering, the advantages of both approaches can be leveraged while mitigating their respective drawbacks: Lobe group shift matches the mechanical orientation. Most lobes retain their intensity, avoiding the attenuation effects seen in pure electronic steering. Because there is no angle mismatch between the electronic and mechanical steering angles, it can overcome the cosine weighting factor from attenuating the $\mathbf{ML}$  intensity and conserves its value.

\begin{figure}[b!]
    \centering
    \includegraphics[width=0.9\columnwidth,clip=true]{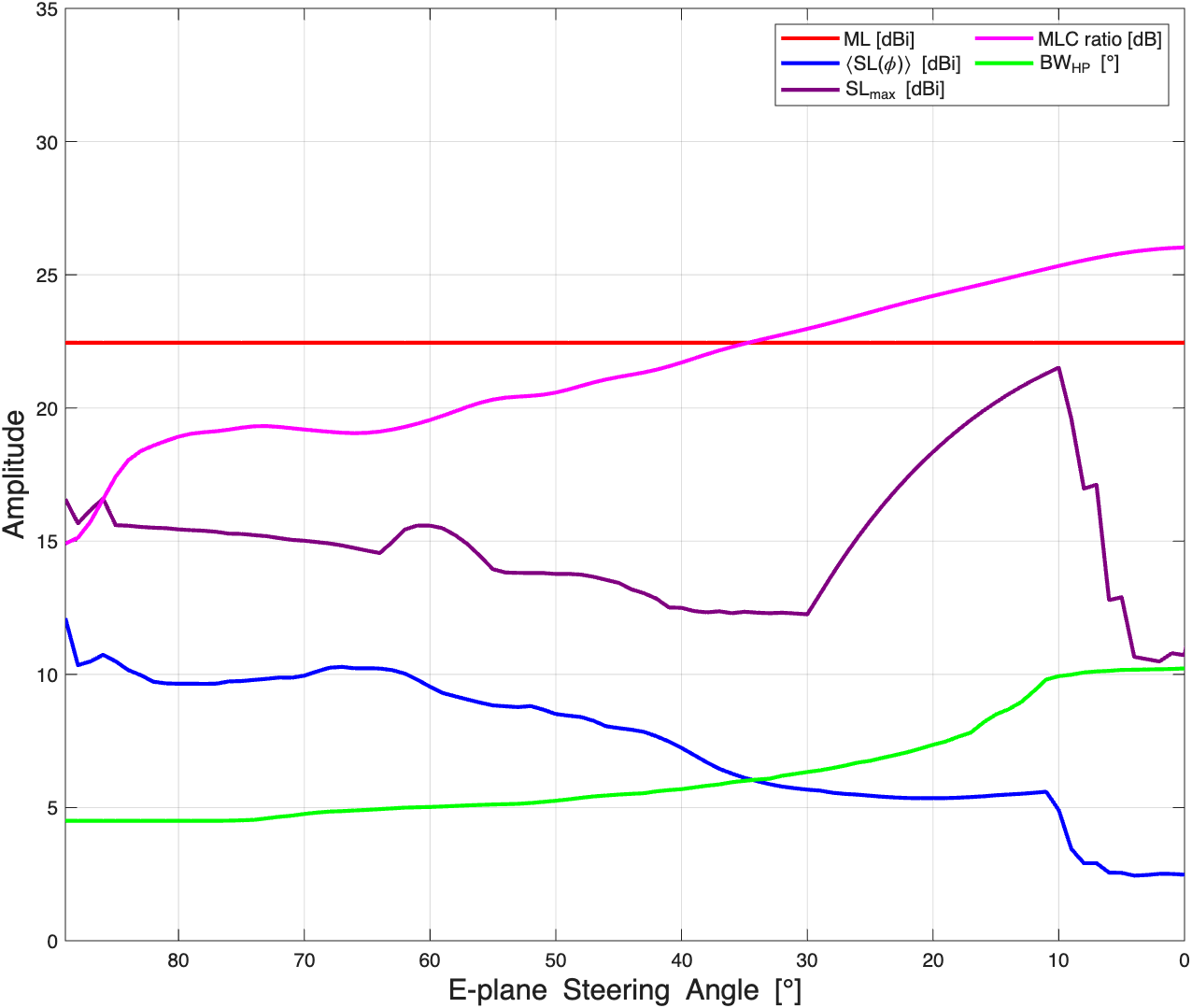}
    \caption{Beam pattern metrics for electronic+mechanical steering sweep for yagi-VHF random phased array.}
    \label{EMsteerThetaSweepVHF}
\end{figure}

Fig.~\ref{EMsteerThetasweep} shows that the $\mathbf{ML}$ remains constant as the E-plane electronic and mechanical steering angle lowers from zenith. For $\mathbf{SL(\phi_0)_{max}}$, its magnitude remains constant until $15\degree$ when it has a huge increase, followed by a decrease below $10\degree$. This is due to a null sector at that angle and when any lobe but the $\mathbf{ML}$ shifts there it looses. For $\mathbf{\langle SL(\phi) \rangle}$ its magnitude decreases. The $\mathbf{MLC}$ ratio shows a constant growth, indicating a $50-1,000$x $\mathbf{ML}$-to-$\mathbf{\langle BP \rangle}$ multiplier through the sweep, indicating that $\mathbf{ML}$ conservation improves the $\mathbf{MLC}$ ratio at low angles due to $\mathbf{\langle SL(\phi) \rangle}$ suppression. The $\mathbf{BW_{HP}}$ remains roughly constant but increases slightly below 10$\degree$, again due to its overlap with the mirror-symmetric back lobe. 

For the VHF array, Fig.~\ref{EMsteerThetaSweepVHF}, we observe similar effects. The main lobe amplitude $\mathbf{ML}$ is constant. For $\mathbf{SL(\phi_0)_{max}}$, its magnitude changes minimally until $30\degree$ when it has a huge increase, followed by a decrease below $10\degree$. This is due to a null sector at that angle and when any lobe but the $\mathbf{ML}$ shifts there it looses. $\mathbf{\langle SL(\phi) \rangle}$ decreases from 12 dBi to 3 dBi over the sweep. The $\mathbf{MLC}$ ratio shows a constant growth, indicating a $30-4,000$x $\mathbf{ML}$-to-$\mathbf{\langle BP \rangle}$ multiplier through the sweep, indicating that $\mathbf{ML}$ conservation improves the $\mathbf{MLC}$ ratio at low angles due to $\mathbf{\langle SL(\phi) \rangle}$ suppression. The $\mathbf{BW_{HP}}$ increases from $5\degree$ to $10\degree$ through the sweep.

\subsection{Array Density Analysis}
\begin{figure}[b!]
    \centering
    \includegraphics[width=0.9\columnwidth]{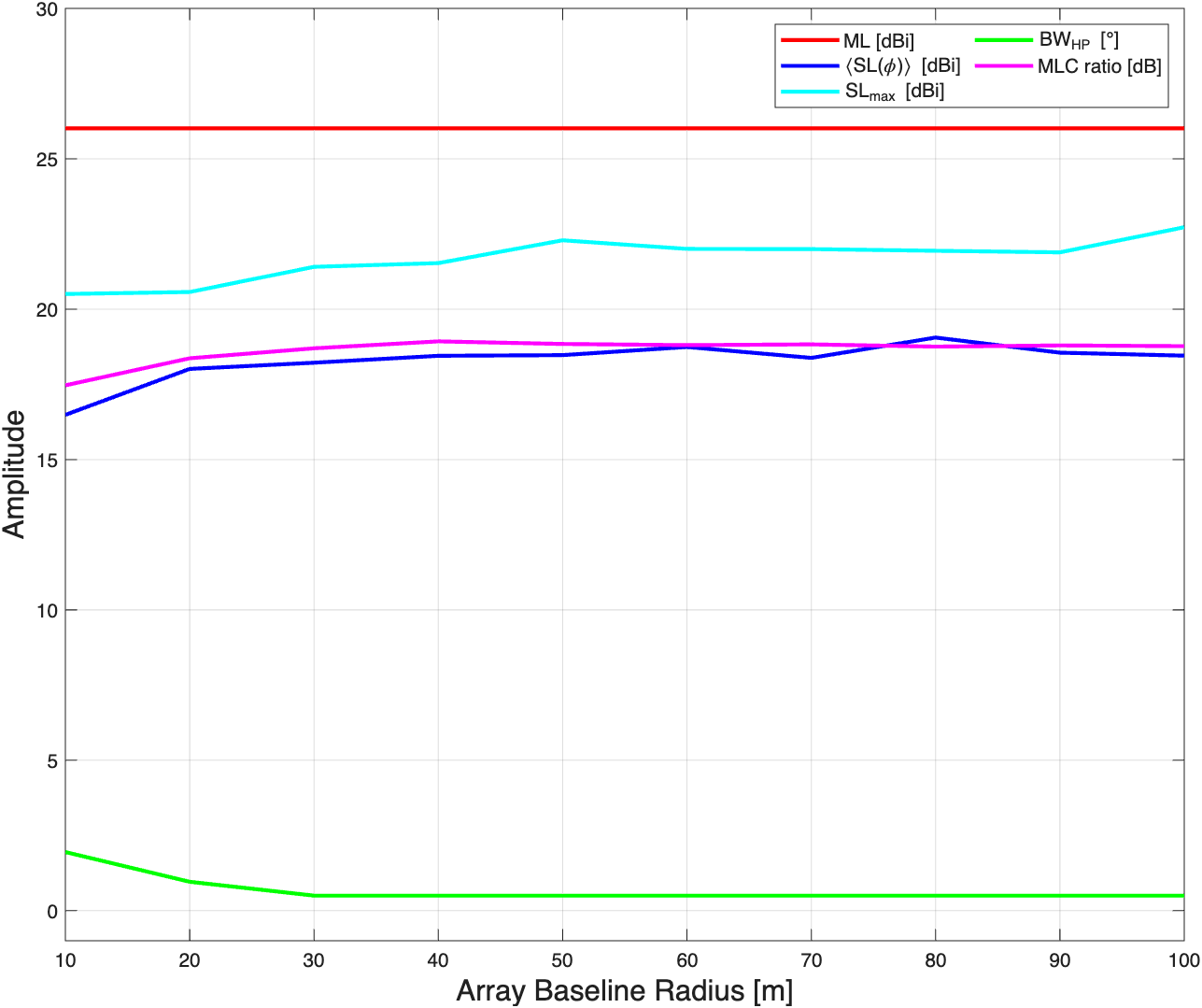}
    \caption{Array density analysis for yagi-UHF phased arrays.}
    \label{ABRA_rand_UHF}
\end{figure}

\begin{figure}[t!]
    \centering
    \includegraphics[width=0.9\columnwidth]{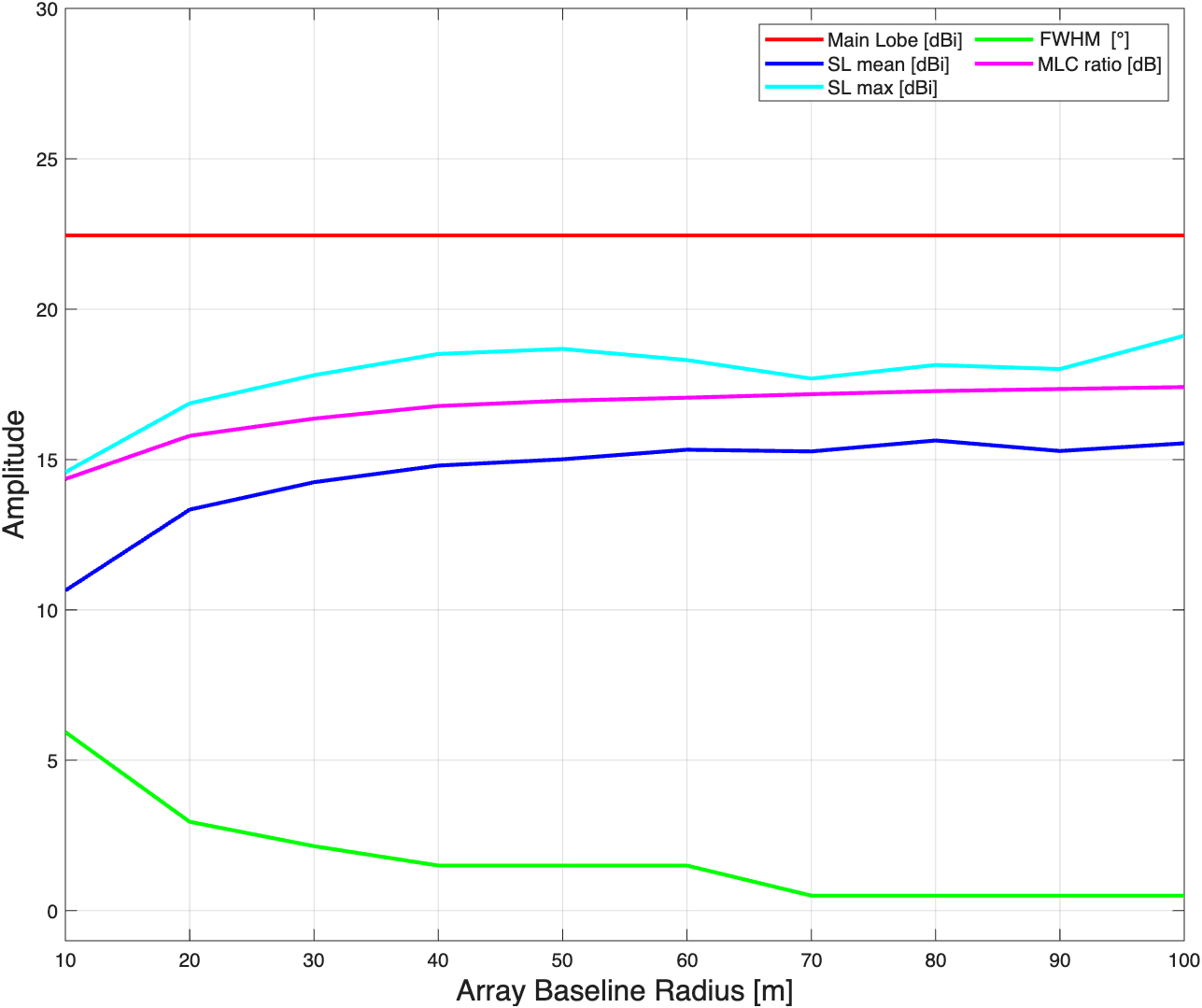}
    \caption{Array density analysis for yagi-VHF phased arrays.}
    \label{ABRA_rand_VHF}
\end{figure}
We will analyze the effects of the inter-element separation distance (3 m to 30 m) and the array's baseline radius (10 m to 100 m) using a 20-element random array. The $\mathbf{ML}$ value ($\sim$26 dBi) remains constant throughout the sweep.

Fig.~\ref{ABRA_rand_UHF} shows the importance of a compact phased array to effectively attenuate side lobes. In the yagi-UHF case, the $\mathbf{\langle SL(\phi) \rangle}$ increases at a low rate, while $\mathbf{SL(\phi_0)_{max}}$ fluctuates slightly but remains within a similar range. The $\mathbf{BW_{HP}}$ decreases steadily up to its minimum at 30 m array baseline radius. This effect is merely an angular sampling effect of the beam pattern, increasing it should allow the $\mathbf{BW_{HP}}$ to keep decreasing. Most importantly, the $\mathbf{MLC}$ ratio remains steady, having a $80$x $\mathbf{ML}$-to-$\mathbf{\langle BP \rangle}$ multiplier.

Fig.~\ref{ABRA_rand_VHF} shows similar behavior for the yagi-VHF configuration. Both $\mathbf{SL(\phi_0)_{max}}$ and $\mathbf{\langle SL(\phi) \rangle}$ increase gradually through the sweep. The $\mathbf{BW_{HP}}$ reaches its minimum at a 70-meter baseline radius. The $\mathbf{MLC}$ ratio remains fairly steady, with a $50$x $\mathbf{ML}$-to-$\mathbf{\langle BP \rangle}$ multiplier.


\section{Conclusion}
The results indicate that a sparse pseudo-random array of yagi-UHF/VHF antennas is a suitable candidate for satellite communications. Compared to a uniform array, it has a high attenuation factor of side lobes from the main lobe.

Our analysis of steering methods revealed that electronic steering shifts all lobes, with main lobe amplitude decreasing with decreasing elevation due to the cosine antenna weight factor, and side lobes increasing as they cross the zenith. 


In combined steering, the field rotates with the steering angles, reducing the cosine weight factor's impact on the main lobe.

We have demonstrated that while maintaining a compact array helps reduce side lobe magnitude, it's also essential to avoid placing antenna elements too closely together, in order to minimize high transmission spectrum cross-talk between antennas. Our findings indicate that reflections occur between antennas at a separation of $0.6\lambda_{UHF/VHF}$, leading to significant cross-talk at these distances. However, these separation distances also contribute to effective side lobe attenuation. Therefore, we conclude that an optimal minimum separation distance between antennas is approximately 3--4 meters.

\section*{Acknowledgment}

We want to thank the South Texas Space Science Institute, the Center for Advanced Radio Astronomy, and the University of Texas Rio Grande Valley for all of the resources, mentoring, and insights that made possible the advancement of this research project.

\bibliographystyle{IEEEtran}
\bibliography{reference}

\begin{IEEEbiography}[{\includegraphics[width=1in,height=1.25in,clip,keepaspectratio]{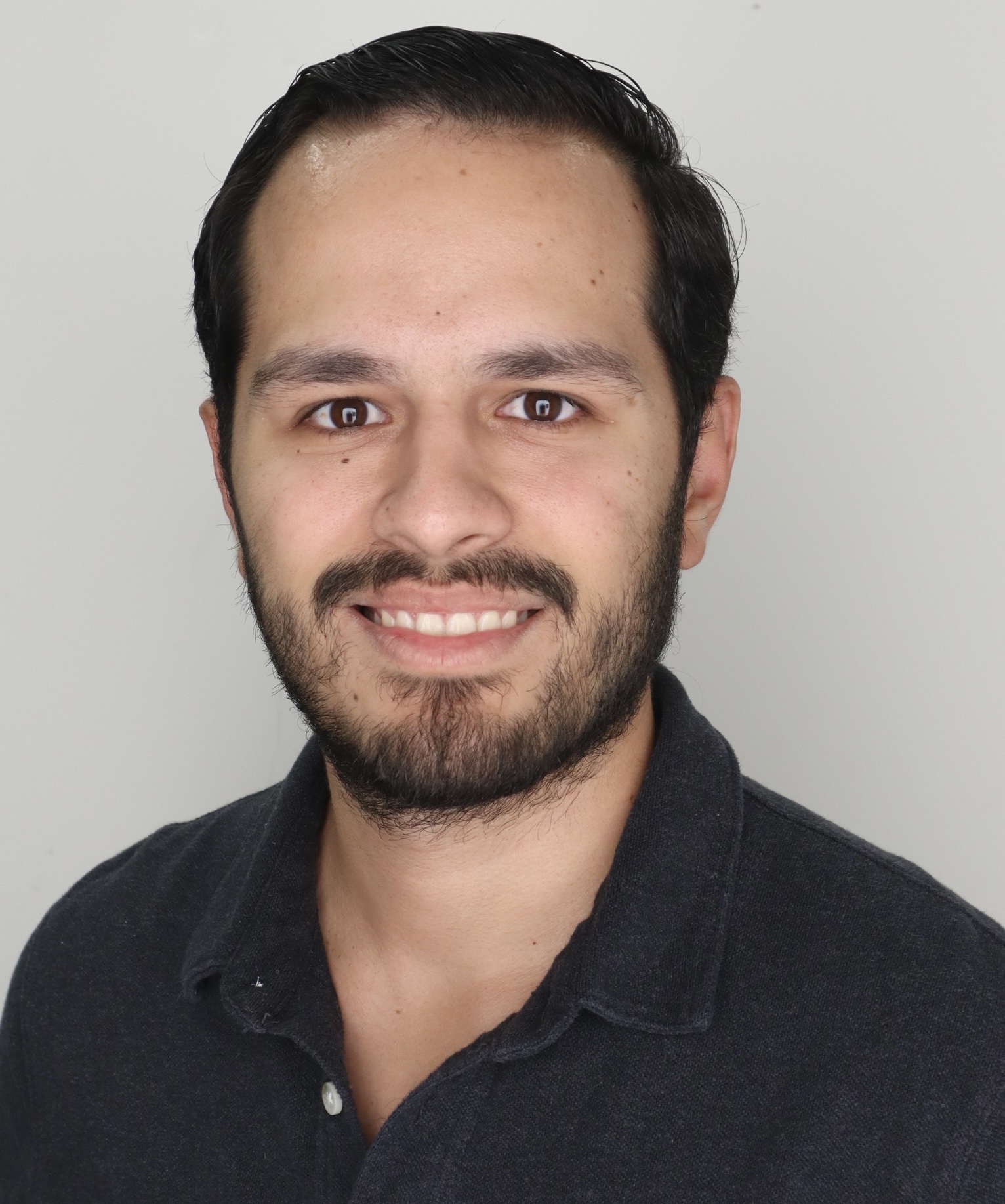}}]{Luis M. Bres} received his PhD. in physics in 2026 at the University of Texas Rio Grande Valley (UTRGV), his M.S. in physics in 2021 at UTRGV, and his B.S. in physics in 2018 at UTRGV.
From 2016-2018 he was an undergraduate research assistant at the Center for Advanced Radio Astronomy (CARA) at UTRGV. From 2019-2021 he was a graduate research assistant at CARA at UTRGV. From 2022-2026 he was a PhD. research assistant at CARA at UTRGV. His research interest include phased array analysis, antenna analysis, systematic evaluation of phased array with different array layouts.
\end{IEEEbiography}

\begin{IEEEbiography}[{\includegraphics[width=1in,height=1.25in,clip,keepaspectratio]{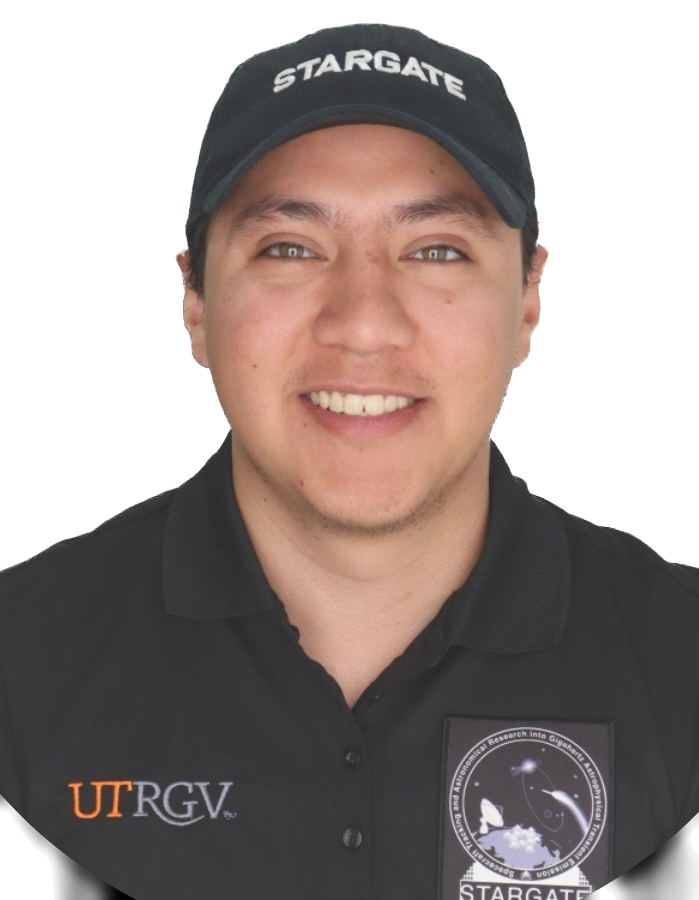}}]{Luis. A., Hernandez} 
received his B.S. degree in mechatronics in 2019 at Tecnologico Nacional de Mexico de Nuevo Laredo, his M.S. in Electrical Engineering in 2023 at UTRGV. Currently he is pursuing his PhD. in Physics at UTRGV. From 2019 to 2022 he was a graduate research assistant on the Blue Energy Project under the guidance of Dr. Yang. From 2022 to date he is a PhD. research assistant at CARA at UTRGV. His research interest are reconfigurable patch antennas, metamaterials analysis, and phased array systems architecture.

\end{IEEEbiography}

\begin{IEEEbiography}[{\includegraphics[width=1in,height=1.25in,clip,keepaspectratio]{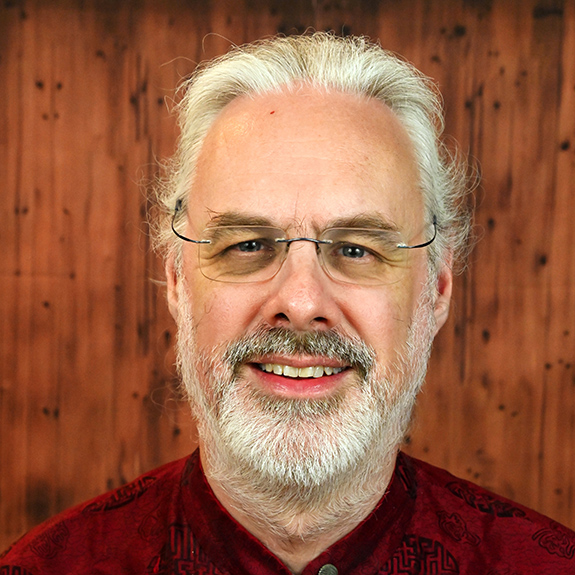}}]{Teviet D. Creighton} received a B.S. degree in physics from the University of Calgary in 1995 and a Ph.D. degree in physics from California Institute of Technology in 2000.

Since 2021, he has been a Professor of Physics and Astronomy at UTRGV, having previously been Associate and Assistant Professor at UTRGV and UT Brownsville (since 2007).  He leads the STARGATE Space Technology Commercialization program at UTRGV's South Texas Space Science Institute.  He is a prolific author of more than 240 articles. His research interests include gravitational wave astrophysics, detection techniques for gravitational waves from compact objects (neutron stars, black holes), and developing Lunar and space-based gravitational-wave detectors.

His distinguished work has been recognized with many awards, including the Special Breakthrough Prize in Fundamental Physics (2016), the Albert Einstein Medal (2017), the Princess of Asturias Award for Technical \& Scientific Research (2017), and the Bruno Rossi Prize (2017).
\end{IEEEbiography}

\EOD

\end{document}